%% file: project_secfed.tex
\documentclass[letterpaper, conference]{IEEEtran}
\IEEEoverridecommandlockouts
\def\BibTeX{{\rm B\kern-.05em{\sc i\kern-.025em b}\kern-.08em
    T\kern-.1667em\lower.7ex\hbox{E}\kern-.125emX}}

\makeatletter
\def\ps@headings{%
\def\@oddhead{\mbox{}\scriptsize\rightmark \hfil \thepage}%
\def\@evenhead{\scriptsize\thepage \hfil \leftmark\mbox{}}%
\def\@oddfoot{}%
\def\@evenfoot{}}
\makeatother
\usepackage[letterpaper]{geometry}
\usepackage{graphicx}
\usepackage{cite}
\usepackage{amsmath,amssymb,amsfonts}
\usepackage{textcomp}
\usepackage{xcolor}
\usepackage[utf8]{inputenc}
\usepackage{bbm}
\PassOptionsToPackage{hyphens}{url}\usepackage[hidelinks]{hyperref}
\usepackage[caption=false]{subfig}
\usepackage{wrapfig}
\usepackage{multirow}
\usepackage{array}
\usepackage[ruled,vlined]{algorithm2e}
\usepackage[inline]{enumitem}
\usepackage{lscape}

\newcommand{\myparagraph}[1]{ \noindent \textbf{#1.}}

\newcommand{\cnr}[1]{{\small\color{red}{CNR: #1}}}

\newcommand{\COne}{\textsc{Comm\_Plain}}

\newcommand{\AggMean}{\textsc{Mean}}
\newcommand{\CTwo}{\textsc{Comm\_Enc}}
\newcommand{\AggEach}{\textsc{Each}}
\newcommand{\CThree}{\textsc{MPC}}
\newcommand{\CVisible}{\textsc{Comm\_Enc\_Limited}}
\newcommand{\PoiM}{\textsc{Poison\_Model}}

\def\extendedVersion{1}  
\newcommand{\ignore}[1]{}

\begin{document}

\title{Network-Level Adversaries in Federated Learning
}

\author{%
  \IEEEauthorblockN{%
    Giorgio Severi\IEEEauthorrefmark{1}\textsuperscript{\textsection},
    Matthew Jagielski\IEEEauthorrefmark{2}\textsuperscript{\textsection}\textsuperscript{\P},
    Gökberk Yar\IEEEauthorrefmark{1},
    Yuxuan Wang\IEEEauthorrefmark{3}\textsuperscript{\P},
    Alina Oprea\IEEEauthorrefmark{1},
    Cristina Nita-Rotaru\IEEEauthorrefmark{1}%
  }%
  \IEEEauthorblockA{\IEEEauthorrefmark{1} Northeastern University, Khoury College of Computer Sciences}%
  \IEEEauthorblockA{\IEEEauthorrefmark{2} Google Research}%
  \IEEEauthorblockA{\IEEEauthorrefmark{3} LinkedIn Corporation}%
}

\maketitle

\begingroup\renewcommand\thefootnote{\textsection}
\footnotetext{These authors contributed equally.}
\endgroup
\begingroup\renewcommand\thefootnote{\P}
\footnotetext{Work performed while being at Northeastern.}
\endgroup

\input{sections/abstract}

\input{sections/intro}

\input{sections/background}
\input{sections/dropping}
\input{sections/defense}
\input{sections/eval_drop}
\input{sections/eval_pois}
\input{sections/eval_visibility}
\input{sections/eval_defense}

\input{sections/related}

\input{sections/conclusion}

\bibliographystyle{IEEEtran}
\bibliography{references,network}

\end{document}

%% file: sections/abstract.tex
\begin{abstract}
Federated learning is a popular strategy for training models on distributed, sensitive data, while preserving data privacy. Prior work identified a range of security threats on federated learning protocols that poison the data or the model. However, federated learning is a networked system where the communication between clients and server plays a  critical role for the learning task performance. We  highlight how communication introduces another vulnerability surface in federated learning and study the impact of network-level adversaries on training federated learning models. We show that attackers dropping the network traffic from carefully selected clients can significantly decrease model accuracy on a target population. Moreover, we show that a coordinated poisoning campaign from a few clients can amplify the dropping attacks. Finally, we develop a server-side defense which mitigates the impact of our attacks by identifying and up-sampling clients likely to positively contribute towards target accuracy. We comprehensively evaluate our attacks and defenses on three datasets, assuming encrypted communication channels and attackers with partial visibility of the network.
\end{abstract}

%% file: sections/intro.tex

\section{Introduction}

Federated Learning (FL) or collaborative learning, introduced by McMahan et al.~\cite{mcmahan_2017_communication}, has become a popular method for training machine learning (ML) models in a distributed fashion.
In FL, a set of clients perform local training on their private datasets and contribute their model parameters to a centralized server.
The server aggregates the local model parameters into the global model, which is then disseminated back to the clients, and the process continues iteratively until convergence. 
Training ML models using FL allows the server to take advantage of a large amount of training data, and each client to retain the privacy of their data.
FL can be deployed in either a cross-silo setting with a small number of participants available in all rounds of the protocol, or in a cross-device setting with a large number of participants, which are sampled by the server and participate infrequently in the protocol. 

Recent work studying the security of FL  has highlighted the risk of attacks by compromised clients through 
data poisoning~\cite{tolpeginDataPoisoningAttacks2020} and 
model poisoning~\cite{bhagojiModelPoisoningAttacks2018,bagdasaryan2020backdoor} under different objectives such as availability, targeted, and backdoor attacks. 
While availability objectives aim at compromising the accuracy of the model indiscriminately~\cite{fangLocalModelPoisoning2020a,Shejwalkar2021ManipulatingTB}, targeted and backdoor attacks only affect a specific subset of samples~\cite{bagdasaryan2020backdoor, sun2019can, wangAttackTailsYes2020} and are harder to detect.
Among the proposed defenses, availability attacks can be mitigated by  gradient filtering~\cite{fangLocalModelPoisoning2020a,Shejwalkar2021ManipulatingTB,signguard2021}, while targeted and backdoor  attacks could be addressed by clipping gradient norms at the server, during model updates, to limit the individual contributions to the global model~\cite{sun2019can,bagdasaryan2020backdoor}. 

FL is a networked system where communication between server and clients plays a critical role
as the global model is learned in rounds with contributions from multiple clients. Thus,
communication represents another point of vulnerability that an attacker can exploit to influence 
the global model learned by the system. In this case, an attacker 
leverages network-level information and attack capabilities to prevent the ML algorithm 
from \emph{receiving} the information needed to learn an accurate model. 
An attacker can exploit the specific communication channels for FL to perturb messages sent between server and clients via jamming~\cite{jamming_cars_vtc_2015}, BGP hijacking~\cite{bgp_hijacking_2019}, compromising routers~\cite{compromised_router_2}, hosts~\cite{overlay_aaron_2008}, or transport-level protocols~\cite{tcpwn_ndss_2018}.  
Network-level attacks have been shown to impact other systems such as  Bitcoin~\cite{Apostolaki2017}, payment-channel networks~\cite{weintraub2020exploiting}, and connected cars~\cite{cacc_matthew_2018}.
These adversaries are especially relevant when political or economic incentives result in a government or company exploiting an AS for which sub-populations are geographically localized. 
For instance, this attack may be used to censor a word prediction model (especially targeting a country's minority language), or modify words associated with a competitor company or unfavorable political movements.

In this paper, we conduct the first study of network-level adversaries' impact on machine learning models trained with federated learning. We analyze the ways an attacker can exploit information sent over the network to maximally damage the learning algorithm under realistic conditions such as encrypted channels and partial visibility of the network. Specifically,
we show that attackers who can carefully orchestrate dropping of the network traffic of selected clients 
can significantly decrease model accuracy on a target population.
Our key insight is that a few clients contribute significantly to the accuracy of the model for target subpopulations, and we create an algorithm that allows the attacker to identify such clients, and, by interfering with only those clients, significantly outperform randomized packet dropping.
Moreover, we show that model poisoning attacks from a small set of clients can further amplify the impact of the targeted dropping attacks. As an example, on a text classification problem, selectively dropping updates from only 5 clients from the target population can lead to a decrease in accuracy from 87\% to 17\%, on a particular target class.
Furthermore, by adding a poisoning attack of equal size, the model accuracy decreases to 0\% on the same target class. 
Finally, we consider a resource limited adversary, who can observe only a small subset of the participating clients, and show that our attacks are still effective, for instance inflicting a 43\% relative accuracy drop when observing only a third of the clients in a computer vision task.

Federated learning system owners do not usually control the underlying network and might not be able to implement network-level mitigations to  make communication more resilient. Complementary to such network-level defenses, we propose a server-level defense that is agnostic to how the dropping attack is performed. 
Our defense modifies client sampling in each round of FL training to increase the likelihood of selecting clients who sent useful updates in previous rounds of the protocol, while decreasing the probability of selecting less relevant clients.
Interestingly, defensive client selection
leverages the same procedure for client identification employed by the network-level adversary. The defense is extremely effective against a targeted dropping attack. For instance, in the same text classification task mentioned above, while an unmitigated attack would completely disrupt the model accuracy on the target task (from 87\% to 17\%), the defense achieves an accuracy of 96\%, which exceeds that of the original model before an attack is mounted. 
Moreover, our defense can be combined with a standard poisoning defense based on update clipping to withstand both targeted dropping and model poisoning attacks. On the same task, the combined dropping and poisoning attack brings the target accuracy to 0, and the combination of our defense with clipping results in 94\% accuracy on the target population. For encrypted communication the improvements compared to the original accuracy can be as high as 34\%.

To summarize, the contributions of the paper are: 
\begin{enumerate*}[label=(\roman*)]
    \item the first study of network-level adversaries in FL, specifically packet dropping attacks targeting a subpopulation and amplification by using model poisoning attacks; 
    
    \item an algorithm for identification of highly contributing clients, allowing the attacker to be more effective than just randomly dropping traffic between clients and server; 
    
    \item a defense based on up-sampling highly contributing clients to the learning task which can be combined with clipping to mitigate both targeted dropping and poisoning attacks;
    
    \item a comprehensive evaluation across multiple model architectures, datasets, and data modalities (image and text).
\end{enumerate*}
For reproducibility, all code is publicly  released\footnote{\url{https://github.com/ClonedOne/Network-Level-Adversaries-in-Federated-Learning}}.

%% file: sections/background.tex

\section{Background and Threat Model}
\label{sec:bk}

\subsection{Federated Learning}
\label{sec:fl}
FL considers a set of $n$ clients, each with a local dataset $D_i$, and a server $S$. 
Clients train locally on their own datasets and 
the server requests model updates, rather than data, building an aggregated global model iteratively over time~\cite{kairouzAdvancesOpenProblems2021}.
We  consider here the  Federated Averaging training algorithm designed for cross-device settings~\cite{mcmahan_2017_communication}. In each round $ 1 \le t \leq T$, the server randomly selects a subset of $m \leq n$ clients to participate in training, and sends them the current global model $f_{t-1}$.
Each selected client $i$ trains locally using dataset $D_i$, for a fixed number of $T_L$ epochs. The server updates the global model $f_t$ by mean aggregation of local updates $U_i$:
$
f_t = f_{t-1} + \eta \frac{ \sum_{i = 1}^{m} U_i } {m}.
$
Data privacy can be further enhanced in FL by secure aggregation performed via Multi-Party Computation (MPC)~\cite{bonawitzPracticalSecureAggregation2017}.

\ignore{\if\extendedVersion1
In this paper, we consider the standard Federated Averaging training algorithm, as described in McMahan et al.~\cite{mcmahan_2017_communication}, shown in Algorithm~\ref{alg:fedlearn}.
\else
We consider the standard Federated Averaging training algorithm, as described in McMahan et al.~\cite{mcmahan_2017_communication}.
\fi
After the global model is initialized to $f_0$ (using the Glorot Uniform initializer~\cite{pmlr-v9-glorot10a}), the training proceeds  for  $T$ rounds.
In each round $ 1 \le t \leq T$, the server randomly selects a subset of $m \leq n$ clients to participate in training, sends them the current global model $f_{t-1}$, and requests model updates from these clients.
Each selected client $i$ trains locally using dataset $D_i$, by performing SGD updates starting from the most recent global model $f_{t-1}$ for a fixed, and usually small, number of $T_L$ epochs.
After the updates $U_i$ are received from the set of participating clients, the server updates the global model $f_t$ by aggregating the client updates:
$$
f_t = f_{t-1} + \eta \frac{ \sum_{i = 1}^{m} U_i } {m}
$$
The updated global model $f_t$ is then broadcast to all the clients in the algorithm at the end of the round. 
}

\if\extendedVersion1
\begin{algorithm}

\small

\KwData{Clients $\mathcal{C}=\lbrace D_i\rbrace_{i=1}^n$, Federated Learning Server $S$, rounds $T$, clients per round $m$, aggregation learning rate $\eta$}

\SetKwFunction{FL}{FedLearn}
\SetKwProg{Fn}{Function}{:}{}
\Fn{\FL{$S, \mathcal{C}$}}{
    \textcolor{gray}{// Function run by server}
    
    $f_0 = \textsc{InitializeModel}()$
    
    \For{$t \in [1, T]$}{
        \textcolor{gray}{// Get updates from $m$ participants in round $i$}
        
        $M_t = \textsc{SelectParticipants}(\mathcal{C}, m)$
        
        $\textsc{RequestUpdate}(f_{t-1}, M_t)$
        
        $U_t = \textsc{ReceiveUpdate}(M_t)$
        
        \textcolor{gray}{// Update and send out new model}
        
        $f_i = \textsc{UpdateModel}(f_{t-1}, U_t, \eta)$
        
        $\textsc{BroadcastModel}(f_t, \mathcal{C})$
    }
}
\caption{Federated Averaging Protocol}
\label{alg:fedlearn}
\end{algorithm}

\fi

\subsection{Threat Model}  
\label{sec:threat}
\noindent {\em Adversarial goal:}
An attacker can target the accuracy for all the classes of the learning task -- availability attacks, or target a particular class -- targeted attacks.  Targeted attacks are much more difficult to detect as the attacker strives to make the model retain its test accuracy for non-targeted points to avoid trivial detection.
We consider targeted attacks and define a population to be one of the classes in the learning task, but this notion could be extended to sub-classes as well. We do not consider poisoning availability attacks which are detectable and can be addressed with existing defenses~\cite{fangLocalModelPoisoning2020a,Shejwalkar2021ManipulatingTB,signguard2021}.

\noindent {\em Attack strategies:} 
The attacker can conduct poisoning attacks, or network-level attack.
Poisoning attacks can target the training data --  an adversary injects maliciously crafted data points to poison the local model~\cite{tolpeginDataPoisoningAttacks2020}, or the model - 
where the adversary compromises a set of clients and sends malicious updates to the protocol with the goal of achieving a certain objective in changing model's prediction~\cite{bagdasaryan2020backdoor, sun2019can, wangAttackTailsYes2020}. Both these attacks are conducted by manipulating directly the inputs to the machine learning algorithm. We consider an
attacker with model poisoning  (\PoiM) capability, using it to amplify network-level attacks. 

In network-level attacks,
based on their network-level capabilities, the adversary can observe communication sent over the network and 
prevent the machine learning algorithm 
from \emph{receiving} the information needed to learn a good model. 
The basic action for a network-level attack is dropping traffic. We consider a smart attacker that selectively drops traffic to maximize the strength of the attack while minimizing detection.
The attacker has several decisions to make: (1) what clients to select to drop their contributions, (2) when to start the attack given that federated learning is an iterative protocol, and (3) how many packets (i.e., local models) to drop. 
We focus on a targeted dropping attack in which the attacker selects a set of clients contributing highly to the target class for dropping their contributions. We compare that with an attack which drops the traffic of a subset of randomly chosen clients.

\noindent \emph{Adversarial network-level knowledge:} 
We consider encrypted communication and partial network visibility 
scenarios for the communication in the FL protocol. For comparison
we also consider unencrypted communication.
\begin{enumerate}
    
   \item \textbf{\COne}. All communication between clients and server is unencrypted. This is a baseline scenario where a network-level adversary obtains maximum information. We do not expect this to happen in practice and we study this model to understand the maximum possible damage.
    
    \item \textbf{\CTwo}. All communication between clients and server is encrypted. This is a realistic scenario, where
     the network-level adversary could infer the set of  clients participating in each round, but not the exact model updates they send. This is the typical deployment for FL, using  IPSec, HTTPS, or application-level encryption.
     
    \item \textbf{\CVisible}. A special case of \CTwo\, where communication is encrypted, and the adversary is limited to only observe a fixed subset of the clients participating in the protocol. This is the most constrained attacker we consider.
    

\end{enumerate}

\noindent {\em Adversarial global model knowledge:}
Since cross-device FL is an open system, the adversary can always participate in the FL protocol as one of the clients. We assume that the adversary obtains the global model updates $f_t$ at each round~$t$.

To summarize, we consider an adversary that has either the \COne, \CTwo\ or \CVisible\ network capability, has knowledge of the global model at each round, and targets a particular victim population of interest. We also consider an adversary that additionally has   \PoiM~poisoning capabilities.

\subsection{Feasibility of Network-level Attacks for Federated Learning}

FL protocols such as  Federated Averaging~\cite{mcmahan_2017_communication} usually abstract the communication protocol between the server and 
clients. In practice, clients will communicate with the server either directly through a 
TCP connection, or through a multi-hop overlay network of hop-by-hop TCP or customized transport 
services. Finally, the last hop connecting the client to the Internet is often in the form of
wireless communication. Network infrastructure and routing protocols facilitate
all this communication through physical or logical connectivity. 
By exploiting the underlying routing protocols and network topology an attacker can position themselves to observe and interfere with data of interest; they can then further impact the accuracy of the global model and create an effect similar to that created through a data
or model poisoning attack. 


Packet dropping attacks can be achieved in multiple ways, and have been studied for network-level adversaries who perform physical-layer attacks in wireless networks~\cite{jamming_cars_vtc_2015}, router compromise~\cite{compromised_router_2}, or transport-level attacks~\cite{tcpwn_ndss_2018}. Each of these methods will require different attacker capabilities. For example for physical layers attacks for wireless clients, the attacker needs to be in their proximity - this is both a powerful attack since jamming is usually difficult to defend against, and limiting since it will be difficult for the attacker to attack  geographically distributed clients without significant resources. In the case of routing, while a router might be more difficult to compromise, in practice the impact 
can be bigger as it might have control over the traffic of multiple clients. Last, but not least, if customized overlays are used
for communication, compromising nodes in the overlay is slightly easier as they are typically hosts, and the number of clients that can be impacted depends on the scalability of the service. 

%% file: sections/dropping.tex

\section{Network-Level Attacks on Federated Learning}
\label{sec:attacks_methodology}

In this section we describe in detail our targeted network attacks against a population in federated learning. 

\subsection{Identification of Highest-Contributing Clients}
The key insight of our attack is that only a small number of clients contribute to the accuracy of a target population of interest. Thus, the most effective and difficult to detect strategy is to drop the traffic just for those highly contributing clients. However, the attacker will first need to identify such clients. 
We design a Client Identification algorithm aimed at 
determining the set of clients whose model updates lead to the largest improvements in target population accuracy.
Our method, described in Algorithm~\ref{alg:client_id}, supports both plain and encrypted network communication, and  is very effective even if the attacker has limited observability of the network.
We demonstrate in Section~\ref{sec:eval_drop} that if the adversary has high success at identifying the clients contributing to the task of interest, then the targeted dropping attack impact is much higher compared to random dropping.

\begin{algorithm}

\scriptsize

\KwData{Target Population Dataset $D^*$, loss function $\ell$, rounds $T_N$, count of clients to drop $k_N$, visibility parameter $v$.}

\SetKwFunction{Diff}{ClientIdentification}
\SetKwProg{Fn}{Function}{:}{}


\Fn{\Diff{$D^*,  \ell, T_N, k_N$}}{
    $f_0 = \textsc{GetGlobalModel}(0)$ 
    
    $\Delta = []$
    
    \For{$t \in [1, T_N]$}{
        
        $M_t=\textsc{GetParticipants}(v)$
        
        $f_t = \textsc{GetGlobalModel}(t)$ 
        
        \If{\CTwo}{
            \textcolor{darkgray}{// Get loss differences for this round}
            
            $L_t = \ell(f_{t-1}, D^*) - \ell(f_t, D^*)$

            \For{$j\in M_t$}{
                \textcolor{darkgray}{// Associate $L_t$ with update members}
                
                $\Delta[j] = \textsc{Concatenate}(\Delta[j], [L_t])$
            }
        }
        \If{\COne}{
            \For{$j\in M_t$}{
                
                $f_t^j = \textsc{GetLocalModel}(j)$
                
                \textcolor{darkgray}{// Get  participant's loss difference}
                
                $L_t^j = \ell(f_{t-1}, D^*) - \ell(f_t^j, D^*)$
                
                \textcolor{darkgray}{// Associate $L_t^j$ with this participant}
                
                $\Delta[j] = \textsc{Concatenate}(\Delta[j], [L_t^j])$
            }
        }

    }
    
    
    
    \textcolor{darkgray}{// Compute the average loss change for each client}
    
    \For{$j \in \Delta$}{
        $\Delta[j] = \frac{1}{|\Delta[j]|}\sum\Delta[j]$
    }
    
    
    $Z = \textsc{GetLargestValues}(\Delta, k_N)$
    
    \Return $Z$
}
\caption{Loss Difference Client Identification}
\label{alg:client_id}
\end{algorithm}

The main intuition is that
the adversary computes the loss of the model before and after the updates on the target class for a number of rounds $T_N$.  Rounds in which the loss decreases correspond to an increase in accuracy on the target class. 
The adversary tracks all the clients participating in those rounds, and computes a loss difference metric per client. 
In the plain communication case, the adversary will determine exactly which clients decrease the loss, while in the encrypted communication case the adversary only observes the aggregated updates and considers all clients participating in the round to be collectively responsible for the loss improvement. 

In more detail, to measure each client's contribution to the target population, the adversary computes the loss difference between successive model updates on the target population (using a representative dataset $D^*$ from the population):
\[
L_{t}^j = \ell(f_{t-1}, D^*) - \ell(f_t^j, D^*)
\]
Note that the second term $\ell(f_t^j, D^*)$  is the loss of the local update of client $j$ in round $t$ for \COne, but becomes the loss of the global model $\ell(f_t, D^*)$ when the adversary only has access to aggregated updates in \CTwo~or \CVisible.
The value $L_t^j$ measures the decrease in the target class loss before and after a given round $t$.

To more reliably estimate client contributions, especially with aggregation of the model updates, the values of $L_{t}^j$ are accumulated over time.
The adversary maintains a list for each client $j$ of the values $L_t^j$ for rounds $t$ in which it participated. 
This allows the adversary to update the list of clients most relevant to the target class accuracy at every round.
Empirically, computing the mean of $L_t^j$ for each client $j$ across all rounds it participates in works well at identifying the clients contributing most to the target class.

\subsection{Dropping Attack}

The targeted dropping attack starts with the adversary performing the Client Identification procedure, by running Algorithm~\ref{alg:client_id} during the first $T_N$ rounds of the protocol, in which  training happens as usual. The adversary identifies in expectation $k_N$ clients contributing updates for the target class. After Client Identification is performed, the network adversary drops contributions from the $k_N$ clients in every round in which they participate after round $T_N$.
As the adversary can expect an increase in identification accuracy by monitoring more rounds of the protocol, they can repeat the Client Identification procedure in each subsequent round, and, if necessary, update the list of selected clients.
We assume that the adversary drops the identified clients' updates, and the server simply updates the model using all received updates. If the Client Identification protocol is not completely accurate, the attacker might drop traffic from 
non-target clients
but we found this has a minimal impact on the trained FL model.




\subsection{Attack Analysis}
\label{sec:analysis}

Algorithm~\ref{alg:client_id} uses several parameters: the number of clients to identify $k_N$ and the number of rounds to wait before identification is successful $T_N$, which we analyze here. 

\subsubsection{How many clients to drop?}
In the FL protocol,  $m$ out of $n$ clients are selected at random in each round. Setting the number of dropping clients $k_N$ is mainly a tradeoff between maximally damaging accuracy on the target data and remaining stealthy by not significantly compromising the overall model performance. If $k_N$ is too large, significant benign updates may be removed, preventing the model from achieving good accuracy, and potentially allowing the server to identify that an active adversary, rather than standard packet loss, is to blame for the dropped updates. If $k_N$ is too small, however, not enough clients will be dropped to have a significant impact on the model. We consider a range of values $k_N \le k$, where $k$ is the number of clients holding examples from the targeted class, and show the attack effectiveness for each.  




\subsubsection{How many rounds are needed to identify the clients?}
Here, we discuss how to set the number of rounds $T_N$ for client identification  for both plain and encrypted communication. When setting $T_N$, if the value is set too small, then the adversary will not have enough observations to reliably identify the contributing clients. However, if set too large, the adversary will allow benign training too much time, resulting in high  accuracy on the target population, making it more difficult to mount the attack later. 
Suppose the adversary wants to wait until all $n$ clients have participated in the protocol at least once. This is a well-studied problem, known as the \emph{coupon collector's}~\cite{CouponCollectorProblem2021} (our setting additionally considers batched arrivals). In this variation of the problem, the adversary must observe an expected number of $cn\log(n)/m$ batches before observing each client, for a small constant $c$. With values of $n=100$, $m=10$, roughly 46 rounds are necessary~\cite{xu2011generalized}. Having established a connection to the coupon collector's problem, we will extend it to model the adversary in each setting:

\myparagraph{Plain communication \COne} 
The baseline case is where the adversary receives \emph{every} client's updates. We carry out a modification of the coupon collector analysis to identify the setting of $T_N$ where a batch of $m$ out of $n$ distinct clients participate in each round, and the goal is to identify a set of $k_N$ of $k$ target clients. 
%
The number of batches to wait for the $i$-th client from the $k$ target clients is $t_i=\tfrac{n}{m(k-i+1)}$ in expectation, as the probability that any target client which has been unobserved appears is $\tfrac{m(k-i+1)}{n}$. By summing this expected batch count over $i$ from 1 to $k_N$ (using the linearity of expectation), we find that the expected number of batches to wait for the first $k_N$ clients is $\tfrac{n}{m}(H_k-H_{k-k_N})$, where $H_i$ is the $i$-th harmonic number (and using $H_0=0$ if $k_N=k$). As $k$ increases, this value tends towards $\approx \tfrac{n}{m}(\ln(k)-\ln(k-k_N))$ rounds (when $k=k_N$, we replace $\ln(0)$ with 0).
To use a setting that is common in our experiments, where $n=60$, $m=10$, $k=15$, and $k_N=15$, we expect to wait roughly $\tfrac{n}{m}\ln(k)=6\ln(15)$ rounds, or roughly 16 rounds. Note that waiting for fewer clients $k_N$ requires fewer rounds.

\myparagraph{Encrypted communication \CTwo~and \CVisible} In the realistic case of encrypted communication, the adversary only has access to mean updates and cannot  localize an improved target accuracy to a particular client, as is possible in \COne.
However, clients who repeatedly participate in rounds where the target accuracy improves, can be considered as more likely targets. Moreover, clients who participate in any round where the target accuracy does not improve can be identified as not targets, and we analyze only rounds without target clients. 
An $\alpha$ precision level at identification can be reached by removing $n-k/\alpha$ clients which are known non-target clients. To analyze how many rounds it takes to collect $n-k/\alpha$ non-target clients, we compute the probability that a batch will contain non-target clients, and then compute how many non-target batches are required to collect $n-k/\alpha$ clients.
To compute the probability that a batch contains non-target clients, we notice that there are $\binom{n-k}{m}$ possible batches which have non-target clients, and there are a total of $\binom{n}{m}$ batches which are selected uniformly at random. Then the probability a batch has non-target clients is $\binom{n-k}{m}/\binom{n}{m}\approx\left(1-\frac{k}{n}\right)^m$.

To compute the number of non-target batches required to accumulate $n-k/\alpha$ non-target clients, we can use comparable coupon collector analysis as before, making the observation that each non-target batch is guaranteed to have $m$ non-target clients. On average, an upper bound of $\tfrac{n}{b}(H_{n-k}-H_{k/\alpha-k})$ batches is sufficient. As an example, in a setting we use in our experiments, $n=60$, $k=15$, and $m=10$, the probability that a batch contains non-target clients is roughly 5.6\%. To reach a precision of $\alpha=0.3$, we obtained a  total of 26 batches on average. In practice, it is likely that precision can be even higher than this value due to overestimation from our analysis.



We note that our analysis is similar to analysis for the \emph{group testing problem}~\cite{GroupTesting2021}, introduced by \cite{dorfman1943detection}, used for pool testing. However, a key difference is that our algorithm must be capable of identifying members from noisy aggregate information, rather than the clean signal which is typically provided during group testing. It is possible that more sophisticated group testing algorithms can be used to improve Algorithm~\ref{alg:client_id} further by overcoming this constraint.


\subsection{Amplifying Dropping Attack with Model Poisoning}

In order to amplify the effectiveness of the targeted dropping attack, the adversary may also collude with, or inject, malicious clients, whose presence in training is designed to further compromise the performance on the target distribution. 
Following Bagdasaryan et al. \cite{bagdasaryan2020backdoor} and Sun et al. \cite{sun2019can}, we use a targeted model poisoning attack known as model replacement. 
Writing $\theta_t$ for the parameter of the global model $f_t$, in this attack, the adversary seeks to replace $\theta_t$ with a selected target $\theta_t^*$ (as is done in \cite{bagdasaryan2020backdoor, sun2019can}, we use the $\theta_t^*$ resulting from a data poisoning attack on a compromised client's local training procedure). The poisoned clients' local training sets are sampled identically to clients with access to the target class, with the difference of changing the labels of target class samples to another, fixed class. 
The update that the adversary sends is $\theta_t^* - \theta_{t-1}$. The server aggregation then weights this update with an $\eta/m$ factor. 
In a model replacement attack, a boosting factor $\beta$ is applied to the update, so the update which is sent is $\beta(\theta_t^* - \theta_{t-1})$, increasing the contribution to overcome the $\eta/m$ factor decrease.  


%% file: sections/defense.tex
\section{Defenses Against Network-Level Adversaries}

Several defenses against network-level attacks are known. 
For instance, defenders could monitor and detect faulty paths in the network~\cite{odsbr_tissesc_2008}, create resilient overlays~\cite{secure_dht_2002, pitn_icdcs_2016}, or secure the routing protocols~\cite{secure_routing_2017}. 
These defenses might increase robustness, but are generally not effective against stealthy attacks, such as targeted update dropping, or edge-level attacks, such as model poisoning. Often, the FL owner might not control the network, so we design FL-specific server defenses that complement existing network-level defenses against the attacks introduced in Section~\ref{sec:attacks_methodology}.

\myparagraph{Model Poisoning Countermeasures} 
A popular form of defense against data and model poisoning attacks in FL is to replace the Federated Averaging protocol with a secure aggregation scheme~\cite{blanchardMachineLearningAdversaries2017,mhamdiHiddenVulnerabilityDistributed2018,yinByzantineRobustDistributedLearning2018a, alistarhByzantineStochasticGradient2018}. It was shown, however, that an adversary can evade this defense (e.g.,~\cite{bagdasaryan2020backdoor,fangLocalModelPoisoning2020a,Shejwalkar2021ManipulatingTB}), and finding a secure aggregation method remains an open problem. 
Sun et al.~\cite{sun2019can} observed that model poisoning attacks with larger gradient norms are more effective, and therefore a natural defense is to reduce the norms. 
With this method, an update $g$ sent from a client is clipped to a maximum norm $C$ and becomes
$\min\left(1, \frac{C}{||\Delta||}\right)g$.
Clipping works particularly well against model poisoning attacks in which the local client update is boosted by the attacker~\cite{bagdasaryan2020backdoor}.

\begin{algorithm}

\scriptsize

\KwData{Target Dataset $D^*_S$, rounds $T_S$, loss function $\ell$, client count $n$, count of clients to upsample $k_S$, up-sample factor $\lambda$}

\SetKwFunction{US}{UpSampling}
\SetKwFunction{Diff}{ClientIdentification}

\SetKwProg{Fn}{Function}{:}{}
\Fn{\US{$D^*_S, \ell, k_S, T_S, n$}}{
    
    $Z =$ \Diff{$D^*_S, \ell, T_S, k_S$}
    
    \textcolor{darkgray}{// Reduce sampling probability for most clients}
    
    $p=[\frac{n - k_S \lambda}{n^2 - k_S n} \text{~for~} c \in [n]]$
    
    \For{$i\in Z$}{
        \textcolor{darkgray}{// Increase sampling probability for highly contributors}
        
        $p[i]=\lambda/n$
    }
    
    \Return $p$
}
\caption{Server Defensive UpSampling Strategy}
\label{alg:upsample}
\end{algorithm}

\myparagraph{UpSampling Defense Strategy} 
While Clipping reduces the impact of model poisoning attack, we still need to defend against targeted update dropping.
When a dropping attack is performed, the number of clients contributing legitimate updates to the target class is reduced, leading to a larger impact on naturally under-represented classes (i.e., available in a small set of clients). 
Thus, it is essential for the server to first identify clients contributing to the target class, and second, leverage them to improve the accuracy of the target population. For the first component, we can use directly Algorithm \ref{alg:client_id} for Client Identification, using the knowledge available to the server. In standard FL implementations, servers receive individual model updates from the clients, while in privacy-preserving FL implementations based on MPC~\cite{bonawitzPracticalSecureAggregation2017} servers only receive aggregated updates. For the second task, the server can run Algorithm~\ref{alg:upsample} (UpSampling defense) to modify the client sampling strategy in the FL protocol. With this modification, the sampling probability of identified clients is increased by a factor of $\lambda$ and the sampling probability of all others is decreased.

The server will determine how many clients to identify, $k_S$,  and how many rounds to monitor, $T_S$, using its own target dataset $D^*_S$ to estimate contributions.
As with the attacker, the server will repeat the Client Identification process at each successive round to refine its list of clients to up-sample.
Interestingly, the UpSampling strategy can help even if there is no dropping attack, but there are simply too few clients from some target population on which the model accuracy is low. 

%% file: sections/eval_drop.tex

\section{Experiemental Evaluation}
\label{sec:eval_drop}

\begin{table}
    \scriptsize
    \centering
    \renewcommand{\arraystretch}{1.1}
    \caption{The parameters used in our experiments.}
    \begin{tabular}{|c|c|c|c|}
    \hline
    \textbf{Party} & \textbf{Param} & \textbf{Definition} & \textbf{Setting} \\
    \hline
    Dataset & $n$ & Total Clients & \{100, 60\} \\
    Dataset & $k$ & Clients with Target Class & \{9, 12, 15\} \\
    Dataset & $\alpha_D$ & Dirichlet Distribution Param & 1.0 \\
    Dataset & $\alpha_T$ & Target Class Dataset Fraction & \{0.5, 0.6\} \\
    \hline
    Client & $T_L$ & Local Training Epochs & 2 \\
    Client & $\eta_L$ & Local Learning Rate & \{ 0.1, 0.001 \} \\
    \hline
    Server & $\eta$ & Aggregation Learning Rate & 0.25 \\
    Server & $m$ & Clients per Round & 10 \\
    Server & $T$ & Total Training Rounds & Varied \\
    Server & $T_{S}$ & Rounds before up-sampling & Varied \\
    Server & $k_S$ & Up-sampled Client Count & Varied \\
    Server & $\lambda$ & Up-sampling factor & 2 \\
    Server & $C$ & Aggregation Clipping Norm & 1 \\
    \hline
    Adversary & $T_{N}$ & Rounds before Dropping & Varied \\
    Adversary & $k_N$ & Number of Dropped Clients & Varied \\
    Adversary & $k_P$ & Number of Poisoning Clients & Varied \\
    Adversary & $\beta$ & Poisoning Boost Factor & 10 \\
    \hline
    \end{tabular}
    \label{tab:params}
\vskip -0.15in
\end{table}

Here, we evaluate our attacks, show how model poisoning amplifies targeted update dropping damage, and conclude by looking at the performance of the UpSampling defense.

\subsection{Experiment Setup}
\label{sec:exp_setup}

We use three well known datasets (EMNIST, FashionMNIST, and DBPedia-14), which we adapt to the \emph{non i.i.d.} setting by controlling the class distribution across clients.
In all cases, the target population is represented by samples of an entire class  selected from the dataset. Our datasets have balanced classes, and we use  classes 0, 1, 9 for our evaluation. All reported results are averages of 4 trials, with different randomness seeds. The list of parameters used in our FL protocol is shown in Table~\ref{tab:params}. 

EMNIST~\cite{cohenEMNISTExtendingMNIST2017} is a  handwritten digits recognition dataset with 344K samples.
We use approximately 100K images of numbers, partitioned equally, without overlap, among 100 clients.
To enforce heterogeneity, we allocate samples from the target victim class to $k$ fixed clients and vary $k$. 
For those $k$ clients, we allocate $\alpha_T=50\%$ of the local dataset to be target class points, while the remainder is sampled from the remaining classes according to a Dirichlet distribution with parameter $\alpha_D=1.0$.
For clients without the target class, we sample 100\% of the local training set from a Dirichlet distribution with $\alpha_D=1.0$, following~\cite{hsu2019measuring}.
We train a convolutional model (two 2D convolution layers and two linear layers, with learning rate $\eta_L=0.1$) for 100 rounds, selecting 10 clients at each round, using mean aggregation with a learning rate $\eta = 0.25$.

FashionMNIST~\cite{xiao2017online} is an image classification dataset, consisting of 70K gray-scale images of 10 types of clothing articles. It is more complex than EMNIST, but smaller. Thus, we increase the number of training rounds to $T \in \{200, 300\}$, reduce the number of clients $n$ to 60, and set $\alpha_T$ to 0.6. We set the local dataset size to 400. We use a convolutional model similar to the one used before, and fix all other parameters.

DBPedia-14~\cite{lehmann2015dbpedia} is an NLP text classification dataset consisting of 560K articles from 14 ontological categories, such as "Company", "Animal", "Film".
DBPedia-14 is a relatively complex dataset, so we use the same $k, T$, and $\alpha_T$ as in FashionMNIST, and a local dataset size of 1000.
The model is trained starting from pre-trained GloVe embeddings~\cite{penningtonGloveGlobalVectors2014}, followed by two 1D convolution layers and two linear layers, and optimized with Adam, with $\eta_L=0.001$.

\subsection{Baselines: Perfect Knowledge and Random Dropping}
\label{sec:baselines}
To demonstrate the potential severity of a dropping attack, we evaluate the best possible case, where the adversary has perfect knowledge of a subset of $k_N$ target clients from the beginning of the protocol, and drops every update originating from them throughout training. 
The results in Table~\ref{tab:random_drop} demonstrate the power of update dropping, and provide a baseline to compare our full attack pipeline against.
We also evaluate the effect of an adversary that selects victim clients uniformly at random.

\begin{table}
    \centering
    \scriptsize
    \renewcommand{\arraystretch}{1.15}
    \caption{Baselines. Target population accuracy at rounds $T / 2$ and $T$. $T = 100$ for EMNIST, $T = 200$ for  FashionMNIST and DBPedia. 
    }
    \begin{tabular}{cc|cccc}
    \multirow{2}{*}{\textbf{Dataset}} & \multirow{2}{*}{$\boldsymbol{k}$} & \multicolumn{4}{c}{\textbf{Number of Clients Dropped} $\boldsymbol{k_N}$}\\
    ~ & ~ & 0 & $k/3$ & $2k/3$ & $k$ \\
    \hline
    \multicolumn{6}{c}{\textbf{Perfect Knowledge}} \\
    \hline
    \multirow{3}{*}{EMNIST} & 9 & 0.47/0.66 & 0.21/0.50 & 0.01/0.23 & 0.0/0.0 \\
    ~ & 12 & 0.58/0.78 & 0.36/0.61 & 0.06/0.31 & 0.0/0.0 \\
    ~ & 15 & 0.65/0.80 & 0.48/0.66 & 0.15/0.40 & 0.0/0.0 \\
    \hline
    FashionMNIST & 15 & 0.40/0.55 & 0.17/0.32 & 0.02/0.09 & 0.0/0.0 \\
    \hline
    DBPedia & 15 & 0.36/0.53 & 0.03/0.06 & 0.00/0.00 & 0.0/0.0 \\
    \hline
    \multicolumn{6}{c}{\textbf{Random Drop}} \\
    \hline
    \multirow{3}{*}{EMNIST} & 9 & 0.47/0.66 & 0.40/0.68 & 0.39/0.68 & 0.35/0.64 \\
    ~ & 12 & 0.58/0.78 & 0.58/0.78 & 0.57/0.77 & 0.53/0.76 \\
    ~ & 15 & 0.65/0.80 & 0.66/0.82 & 0.64/0.81 & 0.65/0.82 \\
    \hline
    FashionMNIST & 15 & 0.40/0.55 & 0.39/0.53 & 0.38/0.53 & 0.35/0.50 \\
    \hline
    DBPedia & 15 & 0.36/0.53 & 0.39/0.54 & 0.31/0.47 & 0.34/0.45 \\
    \hline
    
    \end{tabular}
    \label{tab:random_drop}
\vskip -0.15in
\end{table}

\subsection{Client Identification Evaluation}
\label{sec:cli_identification}

Table~\ref{tab:zero_id_emnist_combined} shows the average number of targeted  clients correctly identified by Algorithm~\ref{alg:client_id} for encrypted  communication with plain as comparison. 
Client Identification works very well for plain communication, and maintains a reasonable performance even when the adversary only sees aggregated updates (\CTwo). For instance, on DBPedia an average of 11.75 out of 15 clients are identified at 70 rounds under \CTwo.
Interestingly, for FashionMNIST and DBPedia, it is easier to identify target clients than for EMNIST (where an average of 7 out of 15 clients are identified at round 70). One hypothesis for this phenomenon is that  classes are more distinct in complex datasets, leading the global model to forget the target class in rounds with no participating target clients, resulting in significant loss increases for those rounds.

\if\extendedVersion1
\begin{table*}
    \centering
    \renewcommand{\arraystretch}{1.2}
    \caption{Target client identification. Average number of clients correctly identified by Algorithm~\ref{alg:client_id} at different rounds  under \COne\ and \CTwo, $k_N=k$. On FashionMNIST and DBPedia all 15 target clients are identified at 50 and 20 rounds, respectively, for \COne, while the maximum number of clients identified under \CTwo\ is 11.75 at 70 rounds for DBPedia.}
    \begin{tabular}{c c c|cccccccc}
    \multirow{2}{*}{\textbf{Dataset}} & \multirow{2}{*}{\textbf{Communication}} & \multirow{2}{*}{$\boldsymbol{k}$} & \multicolumn{8}{c}{\textbf{Round Number $\boldsymbol{T_N}$}}\\
    ~ & ~ & ~ & 5 & 10 & 15 & 20 & 30 & 40 & 50 & 70 \\
    \hline
    \multirow{6}{*}{EMNIST} & \multirow{3}{*}{\COne} & 9  & 1.75 & 5 & 7.50 & 7.75 &  8.50  & 8.75 & 8.75 & 8.50  \\
   
    ~ & ~ & 12 & 5 & 8.25 & 9.75 & 10.25 &10.75 & 10.75 & 11.25 & 11.25 \\
    ~ & ~ & 15 & 4.25 & 9.50  &11.50 & 12  & 14.25 &14.25 &14  &   14   \\
    
    \cline{4-11}
    ~ & \multirow{3}{*}{\CTwo} & 9 & 0.50 & 2.25 &  3.25 & 2.75 & 3.25 & 3.75 & 4 & 5.25 \\
    ~ & ~ & 12 & 2 &  2.75 &3.25 &3 &  3&   4&   5 &5.25 \\
    ~ & ~ & 15 & 3&   4&   4&   3.75 &4.75 &5.75& 6.25 & 7    \\  
    \hline
    \multirow{2}{*}{FashionMNIST} & \COne & \multirow{2}{*}{15} & 9 &  12  & 14 &  14.75 &14.75& 14.75 &15  & 15   \\
    \cline{4-11}
    ~ & \CTwo & ~ & 5.5 & 6.50 & 8.0 & 8.50 & 9 & 10 &  10.50 & 11 \\
    \hline
    \multirow{2}{*}{DBPedia} & \COne & \multirow{2}{*}{15} & 8 &   13.25 &13.75& 15&   15&   15&   15&    15  \\
    \cline{4-11}
    ~ & \CTwo & ~ & 5.25 & 7 &    8&   9&   10&   11.25& 11.25&  11.75   \\
    \hline
    
    \end{tabular}
    \label{tab:zero_id_emnist_combined}
\end{table*}

\else

\begin{table}
    \centering
    \scriptsize
    \renewcommand{\arraystretch}{1.1}
    \caption{
    Average number of target clients identified by Algorithm~\ref{alg:client_id} at different rounds  under \COne\ and \CTwo, $k_N=k$.
    }
    
    \begin{tabular}{c c c|ccccc}
    \multirow{2}{*}{\textbf{Dataset}} & \multirow{2}{*}{\textbf{Comm}} & \multirow{2}{*}{$\boldsymbol{k}$} & \multicolumn{5}{c}{\textbf{Round Number $\boldsymbol{T_N}$}}\\
    ~ & ~ & ~ & 10 & 15 & 30 & 50 & 70 \\
    \hline
    \multirow{6}{*}{EMNIST} & \multirow{3}{*}{\textsc{Plain}} 
          & 9  & 5    & 7.50  &  8.50 & 8.75  & 8.50  \\
    ~ & ~ & 12 & 8.25 & 9.75  & 10.75 & 11.25 & 11.25 \\
    ~ & ~ & 15 & 9.50 & 11.50 & 14.25 & 14    &   14   \\
    \cline{4-8}
    ~ & \multirow{3}{*}{\textsc{Enc}} 
          & 9  & 2.25 & 3.25 & 3.25 & 4    & 5.25 \\
    ~ & ~ & 12 & 2.75 & 3.25 &  3   &   5  & 5.25 \\
    ~ & ~ & 15 & 4    &   4  & 4.75 & 6.25 & 7    \\ 
    \hline
    \multirow{2}{*}{Fashion} & \textsc{Plain} & \multirow{2}{*}{15} 
    &  12  & 14 & 14.75 & 15 & 15 \\
    \cline{4-8}
    ~ & \textsc{Enc} & ~ & 6.50 & 8.0 & 9 & 10.50 & 11 \\
    \hline
    \multirow{2}{*}{DBPedia} & \textsc{Plain} & \multirow{2}{*}{15} 
    & 13.25 & 13.75 & 15 & 15 & 15 \\
    \cline{4-8}
    ~ & \textsc{Enc} & ~ & 7 & 8 & 10 & 11.25 & 11.75 \\
    \hline
    
    \end{tabular}
    \label{tab:zero_id_emnist_combined}
\vskip -0.15in
\end{table}

\fi

We also use Table~\ref{tab:zero_id_emnist_combined} to select parameters for the targeted dropping attack and validate our analysis from Section~\ref{sec:analysis}.  
In the \CTwo{} scenario, we see that more rounds are necessary for successful identification, as expected from Section~\ref{sec:analysis}. To identify between $k/3$ and $2k/3$ of the target clients, we need to wait between 30 and 50 rounds, and, in many cases, the identification accuracy tends to plateau in successive rounds.
Thus, we select round 30 as the starting point for the dropping attack in our \CTwo\ experiments.

\subsection{Targeted Dropping Evaluation}
\label{sec:id_drop}

We measure our targeted dropping attack's performance in Table~\ref{tab:identification_drop}.
Under \COne, it significantly compromise target population accuracy, closely approximating the perfect knowledge adversary for increasing values of $k_N$.
We also observe that our attack in the \COne\ scenario vastly outperforms the strategy of randomly dropping the same number of clients, in all situations.
Our attacks' performance decreases when the adversary's knowledge is limited in \CTwo, which is expected from the reduced Client Identification performance.
We still observe a significant advantage in using our identification pipeline, over the random selection baseline.
For instance, with $k_N=5$, the target accuracy on DBPedia drops drastically from 53\% to 6\%.
Moreover, these results highlight the trend mentioned in Section~\ref{sec:cli_identification}: on more complex tasks, such as FashionMNIST and DBPedia, the high identification accuracy leads to noticeably larger attack performance, than in EMNIST.
Given the targeted nature of our attack, in all considered scenarios, dropping the victim clients generally leads to very little degradation of the overall model performance -- on average 3.88\% for \COne\ and 1\% for \CTwo. 

\if\extendedVersion1
Tables~\ref{tab:upsample_global_each}, \ref{tab:upsample_global_mean}, \ref{tab:upsample_global_c3}, show the accuracy of the global model on the full test set under the same settings as the other experiments we run. 
\fi

\begin{table}
    \centering
    \scriptsize
    \renewcommand{\arraystretch}{1.1}
    \caption{Targeted dropping attack, under \COne\ and \CTwo. Accuracy on target population at rounds $T / 2$ and $T$. $T = 100$ for EMNIST, $T = 200$ for FashionMNIST and DBPedia.
    }
    \begin{tabular}{cc | cccc}
    
    \multicolumn{6}{c}{ \multirow{2}{*}{\textbf{\COne}} } \\
    \\
    \hline
    \multirow{2}{*}{\textbf{Dataset}} & \multirow{2}{*}{$\boldsymbol{k}$} & \multicolumn{4}{c}{\textbf{Number of Clients Dropped} $\boldsymbol{k_N}$}\\
    ~ & ~ & 0 & $k/3$ & $2k/3$ & $k$ \\
    \hline
    
    \multirow{3}{*}{EMNIST} & 9 & 0.47/0.66 & 0.25/0.49 & 0.09/0.18 & 0.00/0.00 \\
    ~ & 12 & 0.58/0.78 & 0.39/0.65 & 0.19/0.33 & 0.00/0.00 \\
    ~ & 15 & 0.65/0.80 & 0.50/0.74 & 0.26/0.50 & 0.02/0.02 \\
    \hline
    FashionMNIST & 15 & 0.40/0.55 & 0.24/0.23 & 0.02/0.03 & 0.00/0.00 \\
    \hline
    DBPedia & 15 & 0.36/0.53 & 0.10/0.01 & 0.00/0.00 & 0.00/0.00 \\

    \hline
    \multicolumn{6}{c}{ \multirow{2}{*}{\textbf{\CTwo}} } \\
    \\
    \hline

    \multirow{3}{*}{EMNIST} & 9 & 0.47/0.66 & 0.35/0.58 & 0.32/0.52 & 0.32/0.52 \\
    ~ & 12 & 0.58/0.78 & 0.50/0.72 & 0.48/0.69 & 0.45/0.67 \\
    ~ & 15 & 0.65/0.80 & 0.63/0.78 & 0.60/0.76 & 0.56/0.71 \\
    \hline
    FashionMNIST & 15 & 0.40/0.55 & 0.34/0.38 & 0.14/0.13 & 0.03/0.01 \\
    \hline
    DBPedia & 15 & 0.36/0.53 & 0.19/0.06 & 0.06/0.00 & 0.00/0.00 \\
    \hline
    
    \end{tabular}
    \label{tab:identification_drop}
\vskip -0.15in
\end{table}

%% file: sections/eval_pois.tex
\begin{figure}
    \centering
    
    \subfloat[Perfect knowledge\label{fig:emnist_pk_acc100}]{%
      \includegraphics[width=0.5\linewidth]{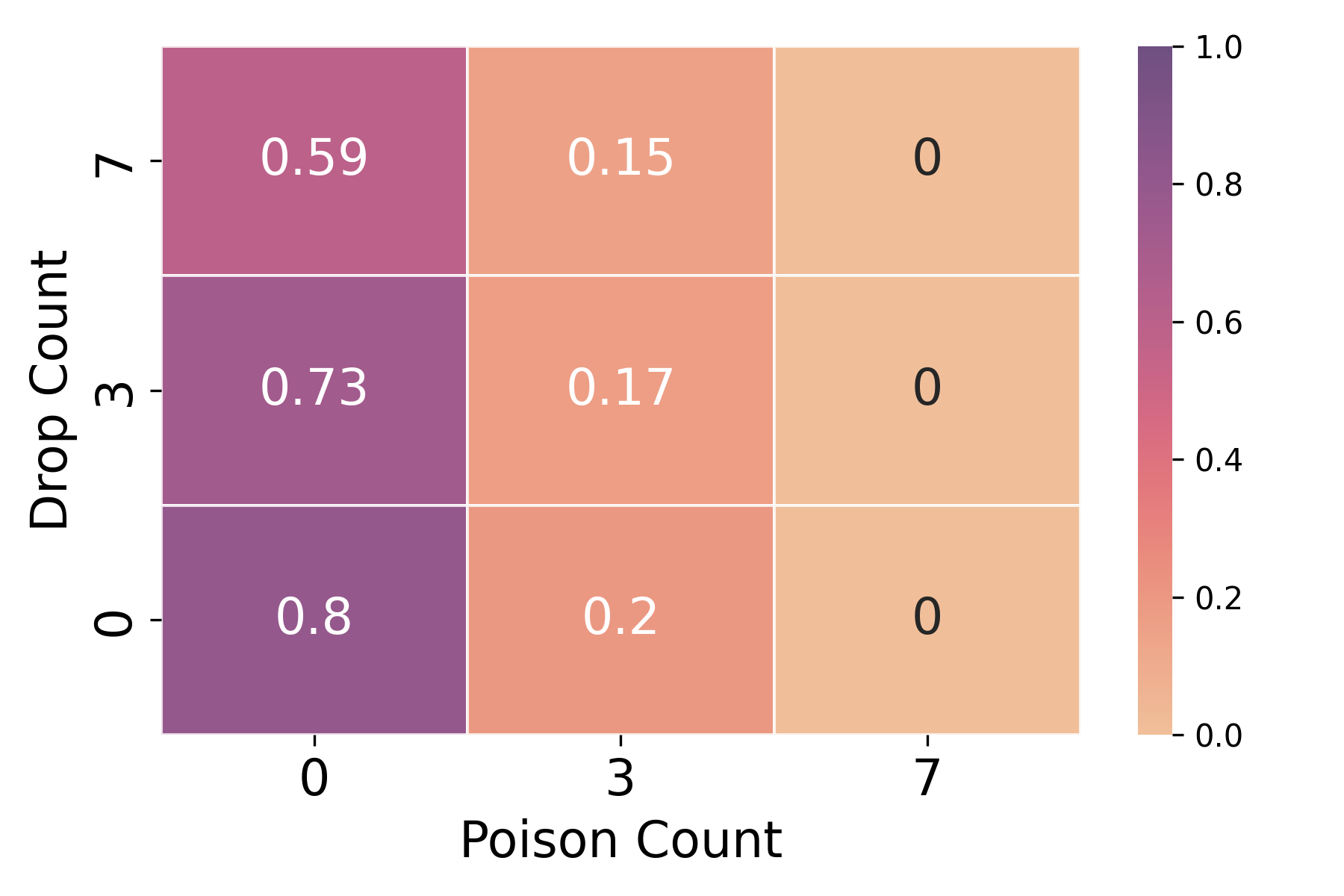}}
    \subfloat[Perfect knowledge, clipping\label{fig:emnist_pk_clip_acc100}]{%
      \includegraphics[width=0.5\linewidth]{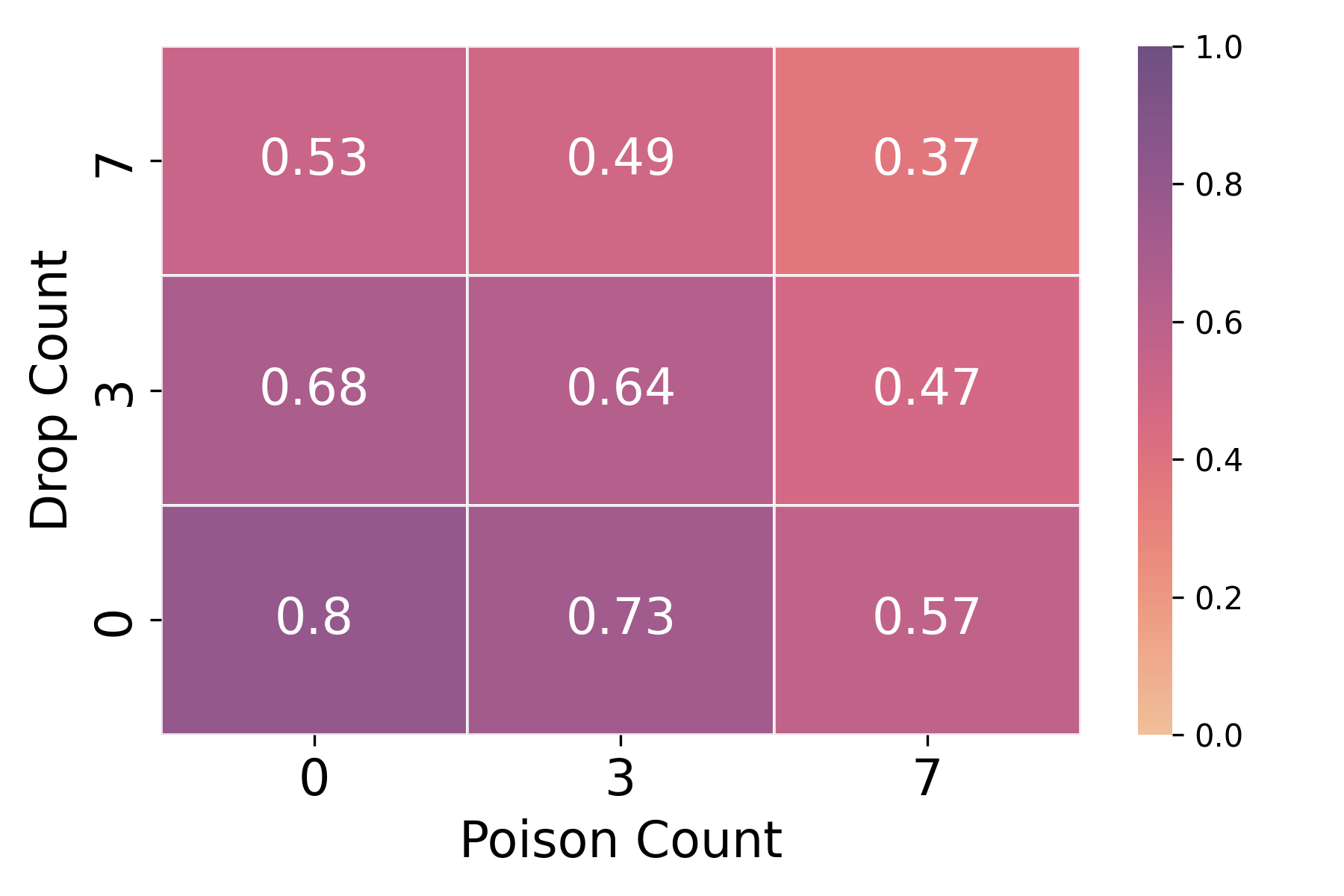}}
    \\
    \vskip -0.1in
    \subfloat[\COne\label{fig:emnist_c1_acc100}]{%
        \includegraphics[width=0.5\linewidth]{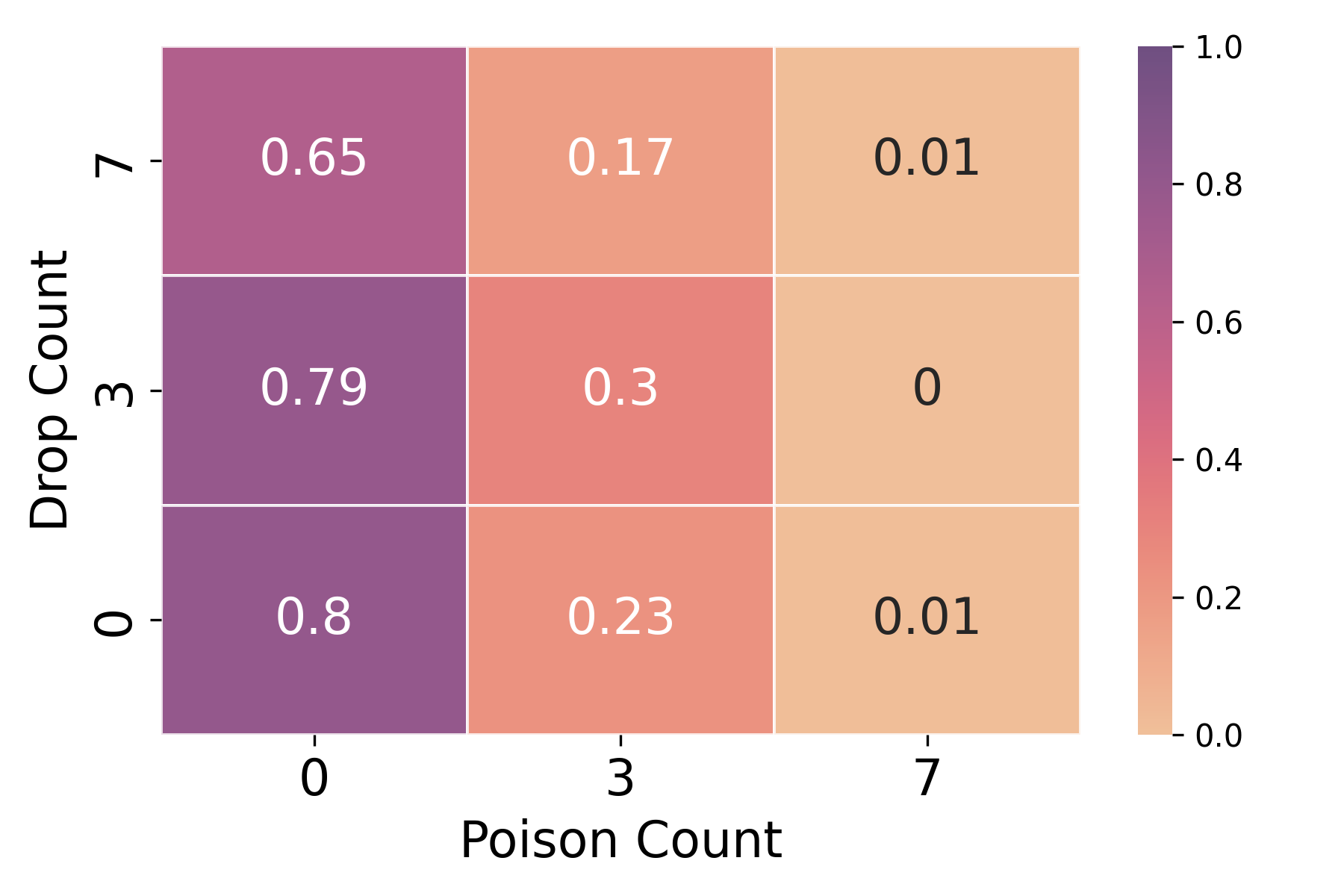}}
    \subfloat[\COne, clipping\label{fig:emnist_c1_clip_acc100}]{%
        \includegraphics[width=0.5\linewidth]{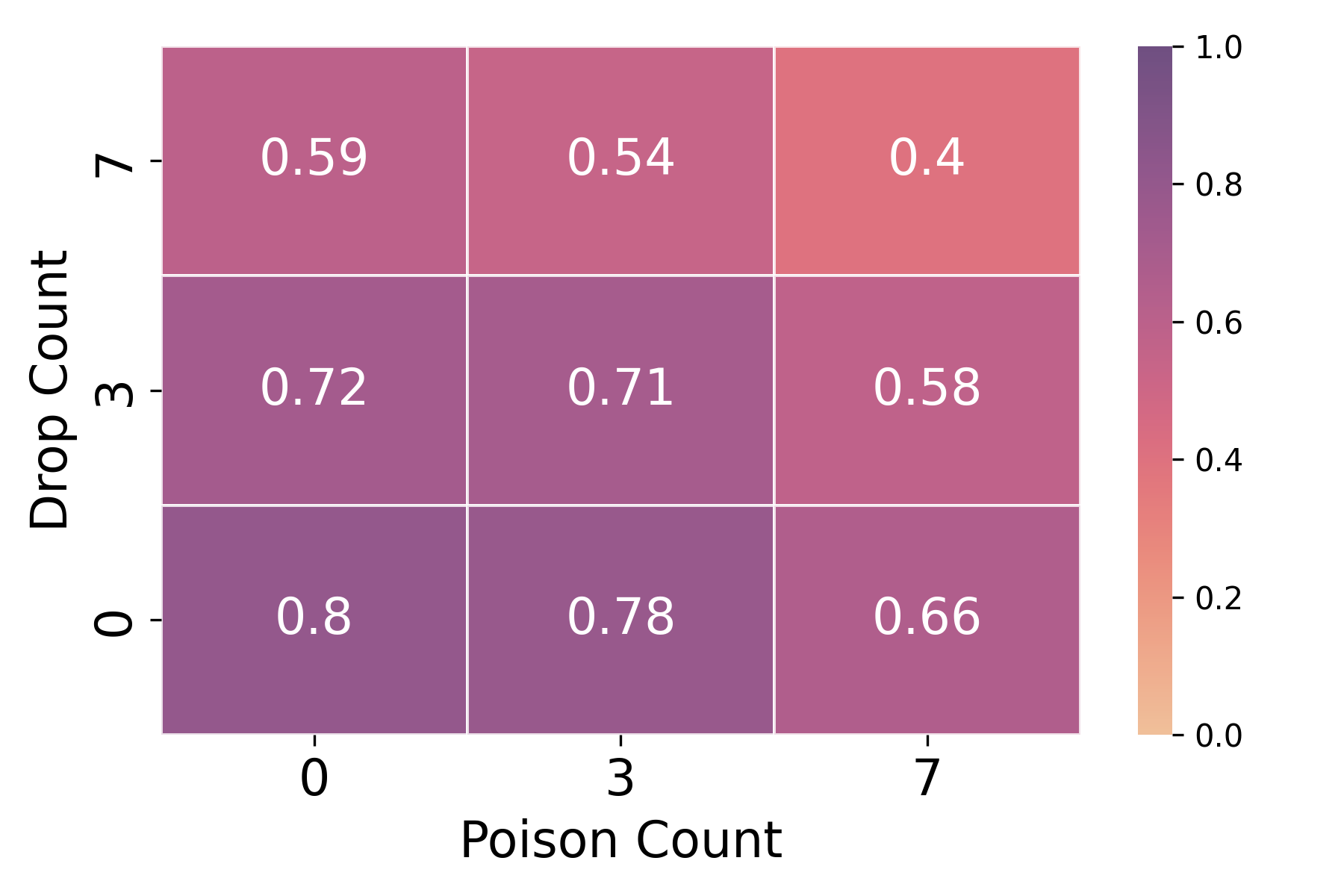}}       
    \\
    \vskip -0.1in
    \subfloat[\CTwo\label{fig:emnist_c2_acc100}]{%
        \includegraphics[width=0.5\linewidth]{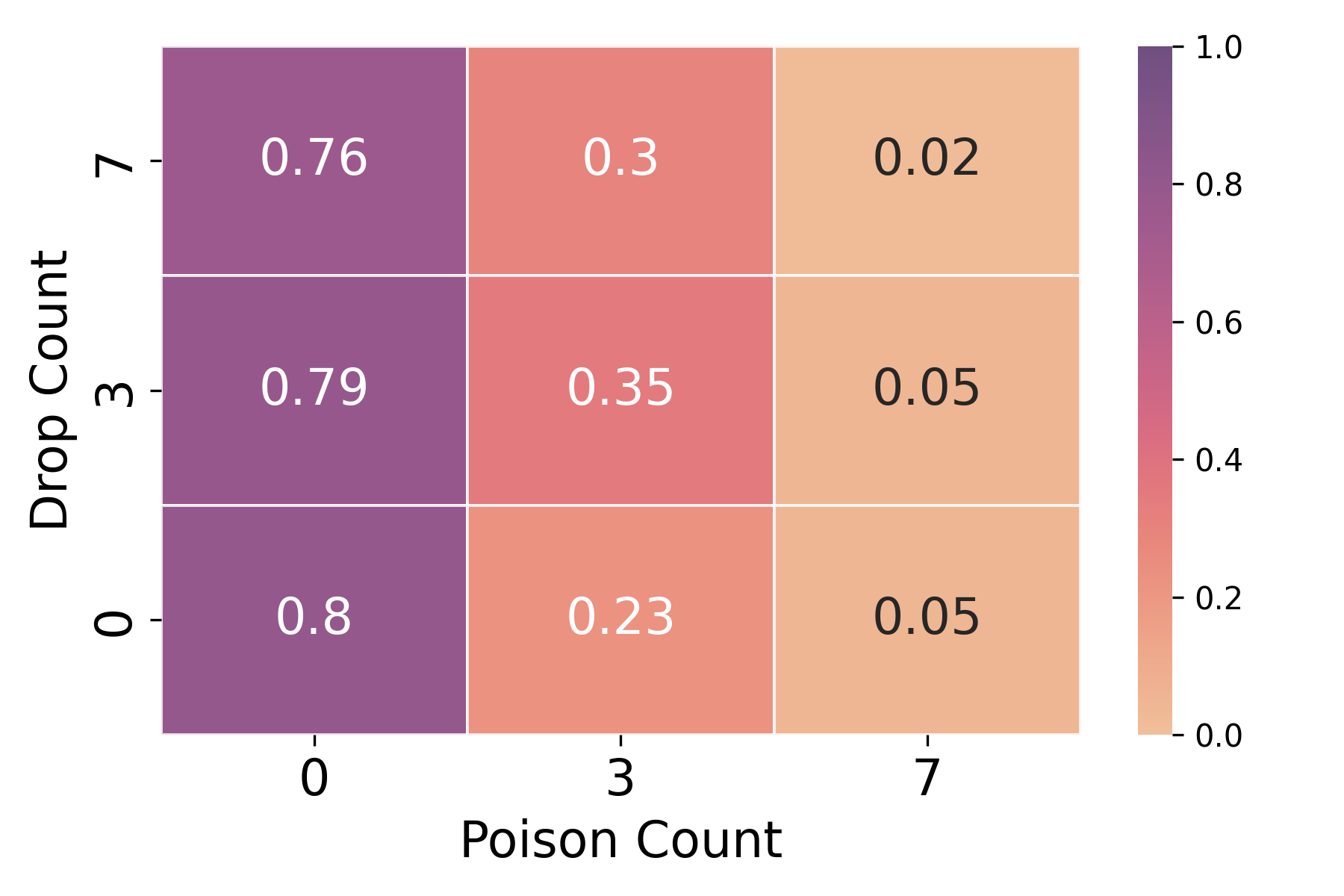}}
    \subfloat[\CTwo, clipping\label{fig:emnist_c2_clip_acc100}]{%
        \includegraphics[width=0.5\linewidth]{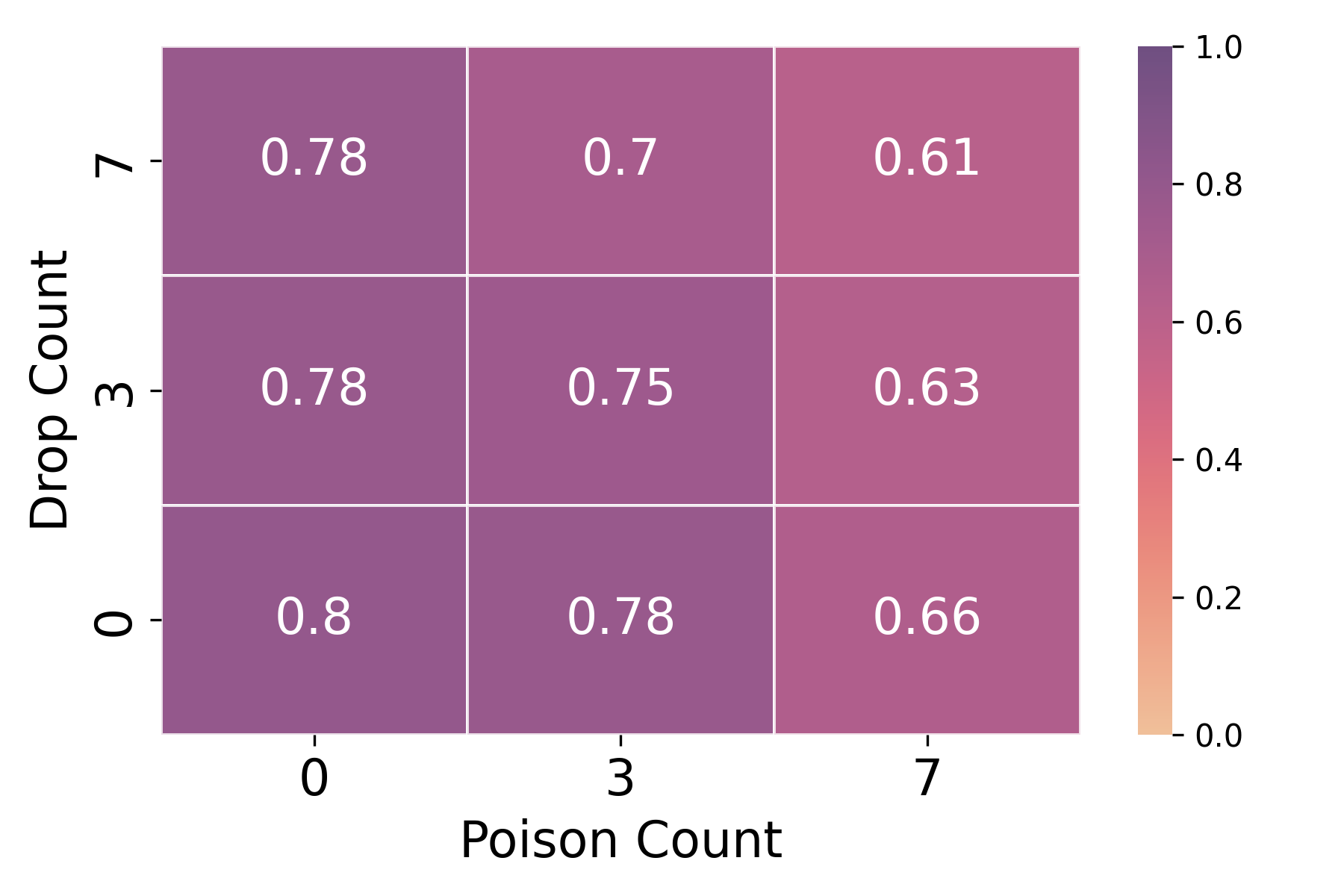}}
    
    \caption{Accuracy on  target class 0 on EMNIST, for $k=15$, $T = 100$, and varying number of dropped and poisoned clients under 3 scenarios: Perfect Knowledge, \COne, and \CTwo. Left results are without Clipping, and right results use a Clipping norm of 1.
    }
    \label{fig:pois_and_drop_clip}
\vskip -0.18in
\end{figure}

\subsection{Impact of Model Poisoning and Targeted Dropping}

We compared the effects of the model poisoning strategy and targeted dropping on the EMNIST dataset for the cases of perfect knowledge, \COne, and \CTwo\ in Figures~\ref{fig:emnist_pk_acc100}, \ref{fig:emnist_c1_acc100}, and \ref{fig:emnist_c2_acc100} respectively.
These heatmaps show that, for different levels of $k_N$ (number of dropped clients) and $k_P$ (number of poisoned clients), the model replacement attack is more effective than targeted dropping, even when the adversary has perfect knowledge.
The results, however, are significantly different when Clipping, a standard poisoning defense~\cite{sun2019can, gupta2020byzantine,hong2020effectiveness}, is applied to local model updates.
Figures~\ref{fig:emnist_pk_clip_acc100}, \ref{fig:emnist_c1_clip_acc100}, and \ref{fig:emnist_c2_clip_acc100} show the same set of experiments repeated with a clipping norm of 1.
These heatmaps highlight that, while clipping lowers the impact of model poisoning, the combination of targeted dropping and model poisoning still results in very noticeable targeted performance degradation. 
For instance, under \COne, the original model accuracy on class 0 is 0.8. 
This becomes 0.59 when dropping 7 clients out of 15, and 0.66 when poisoning 7 clients, while combining both attacks leads to 0.4 accuracy.

\if\extendedVersion1
    
    

\begin{figure}
    \centering
    
    \subfloat[Perfect knowledge\label{fig:emnist_pk_acc50}]{%
       \includegraphics[width=0.5\linewidth]{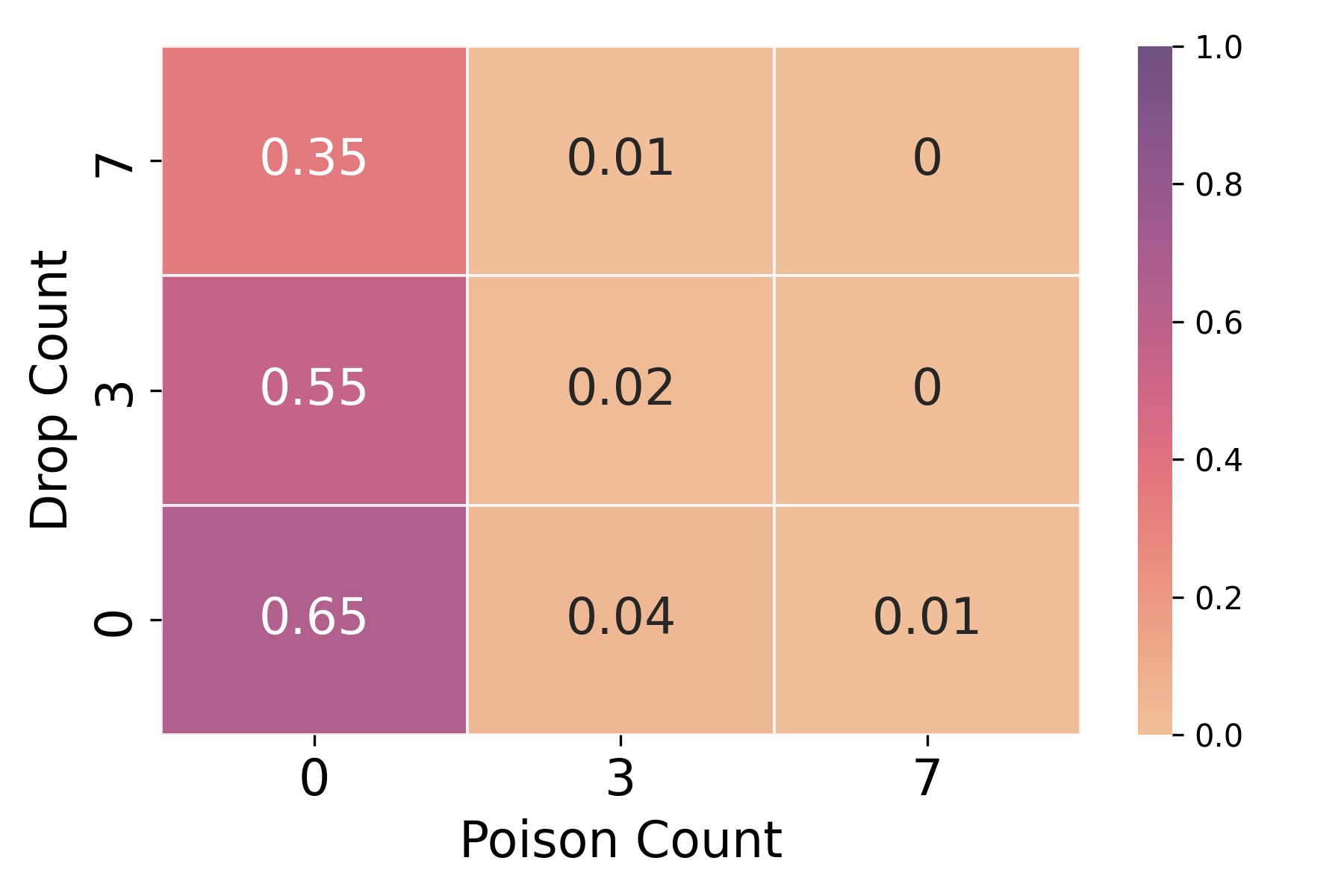}}
    \subfloat[Perfect knowledge, clipping\label{fig:emnist_pk_clip_acc50}]{%
       \includegraphics[width=0.5\linewidth]{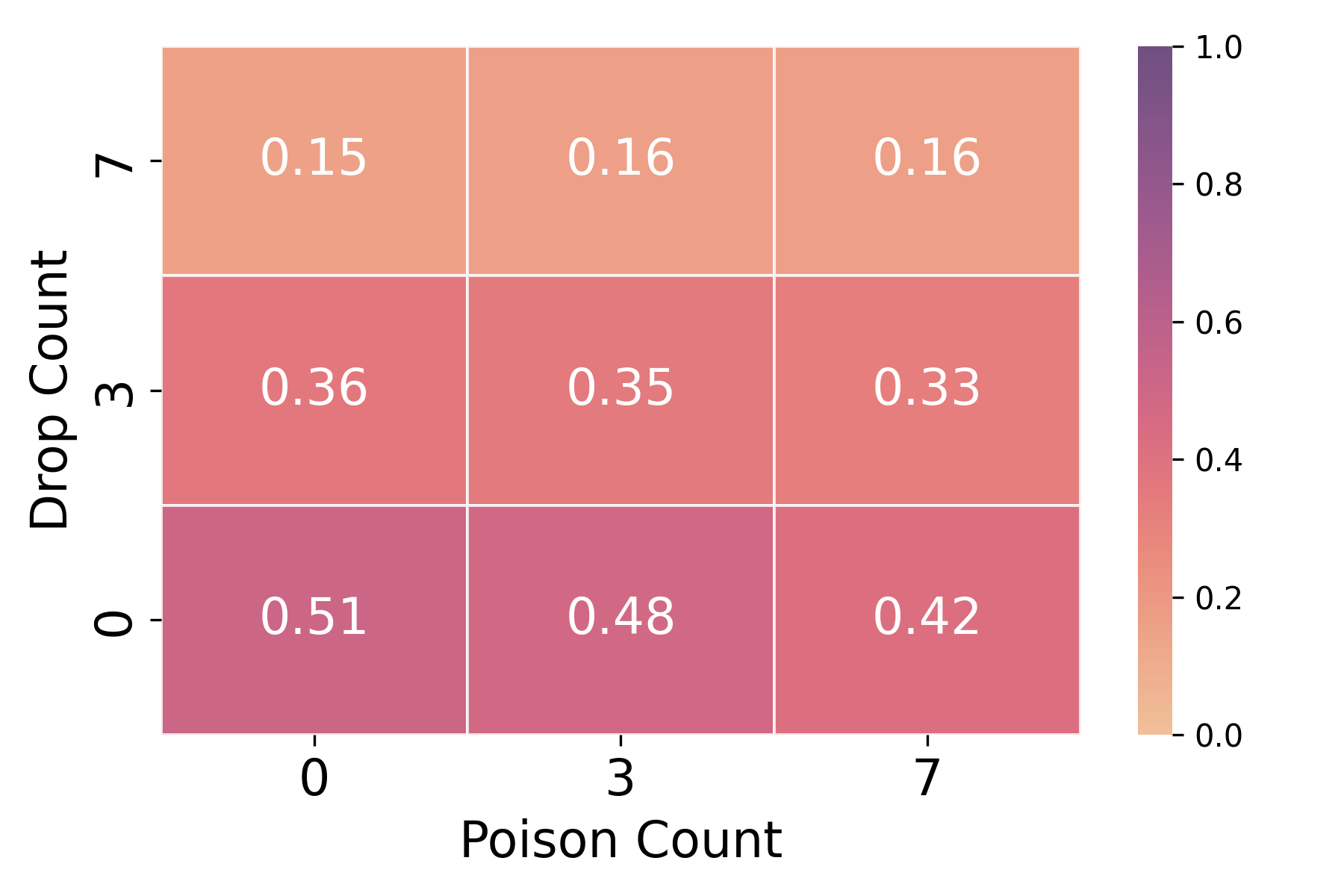}}
    \\
    \subfloat[\COne\label{fig:emnist_c1_acc50}]{%
        \includegraphics[width=0.5\linewidth]{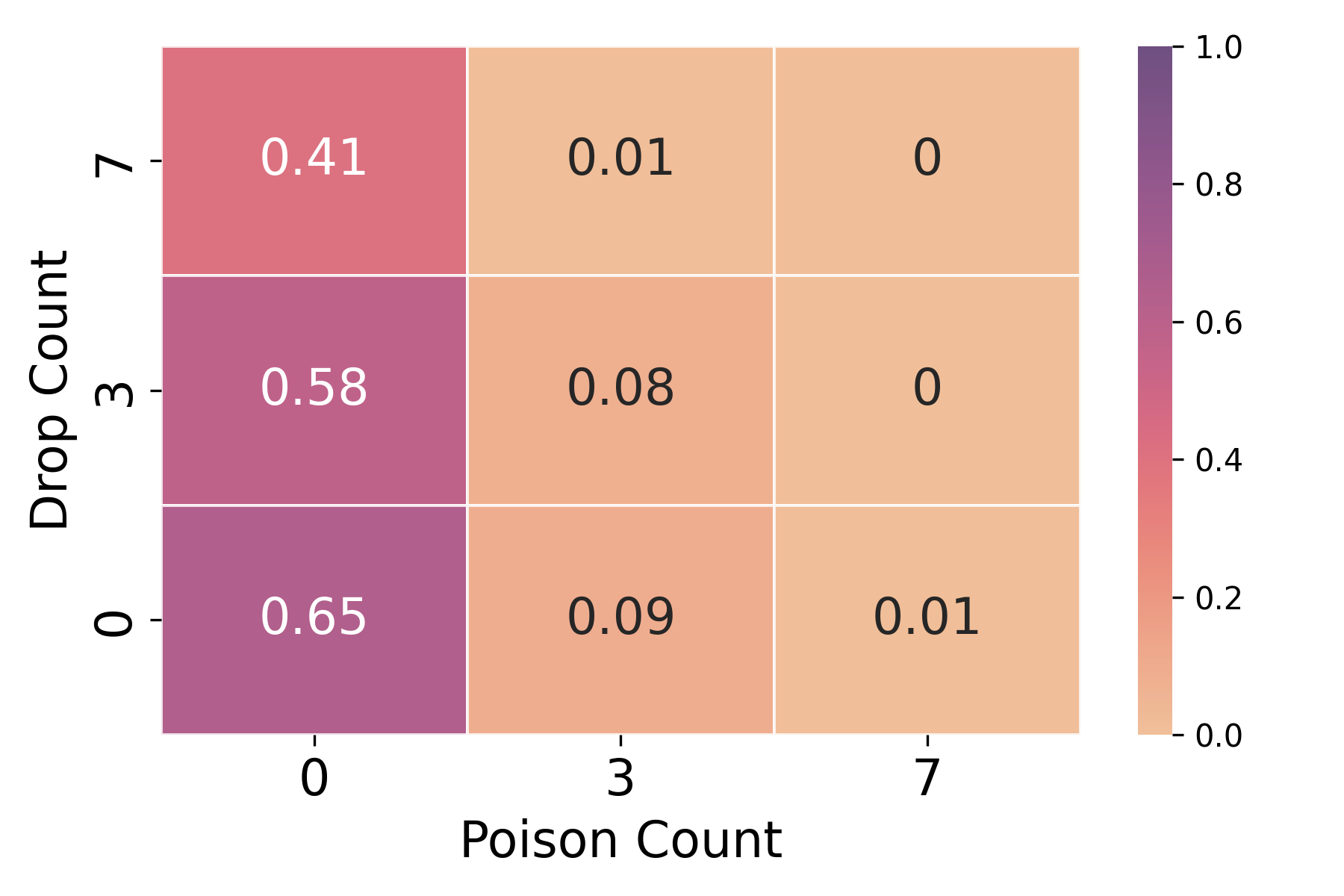}}
    \subfloat[\COne, clipping\label{fig:emnist_c1_clip_acc50}]{%
        \includegraphics[width=0.5\linewidth]{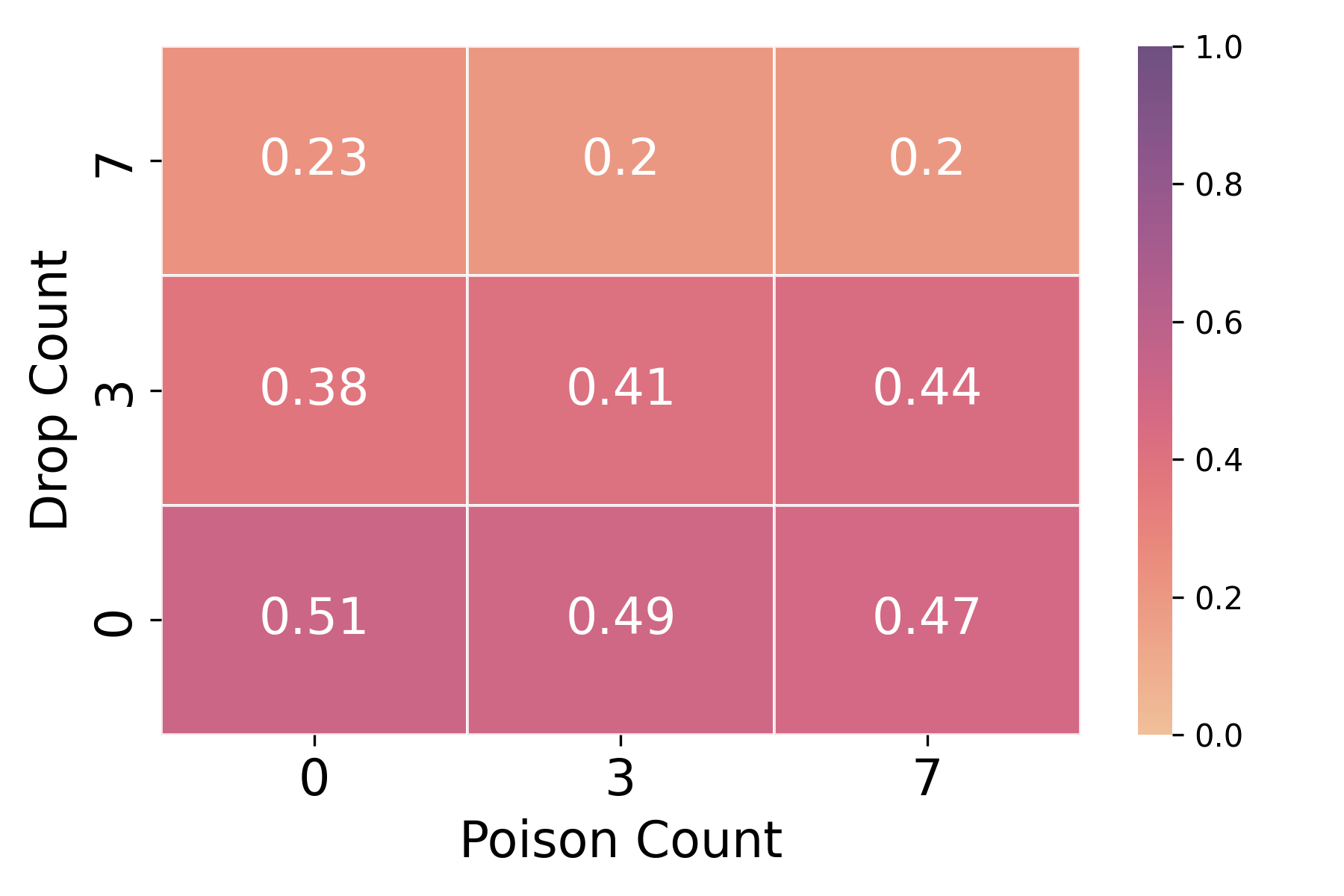}}
    \\
    \subfloat[\CTwo\label{fig:emnist_c2_acc50}]{%
        \includegraphics[width=0.5\linewidth]{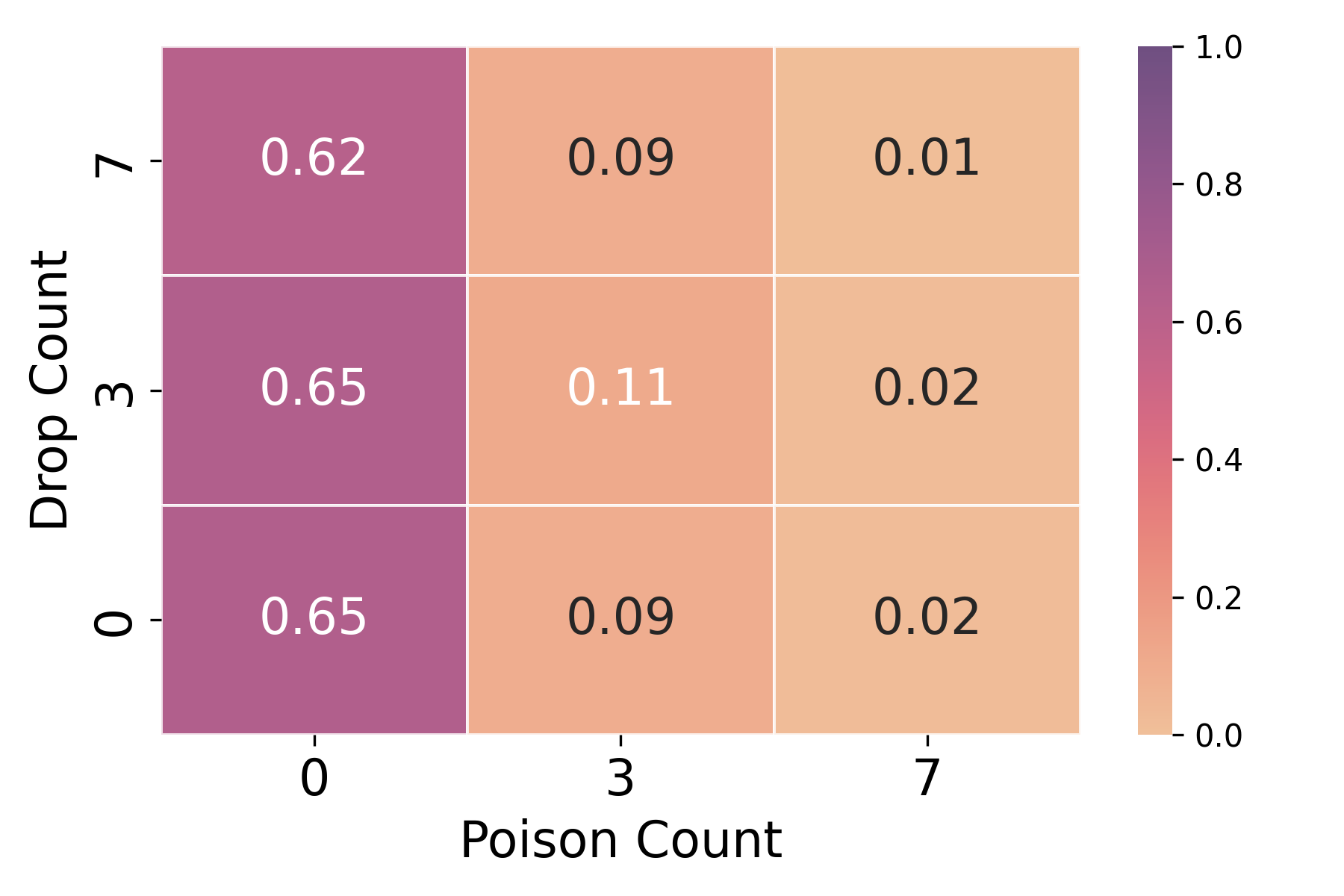}}
    \subfloat[\CTwo, clipping\label{fig:emnist_c2_clip_acc50}]{%
        \includegraphics[width=0.5\linewidth]{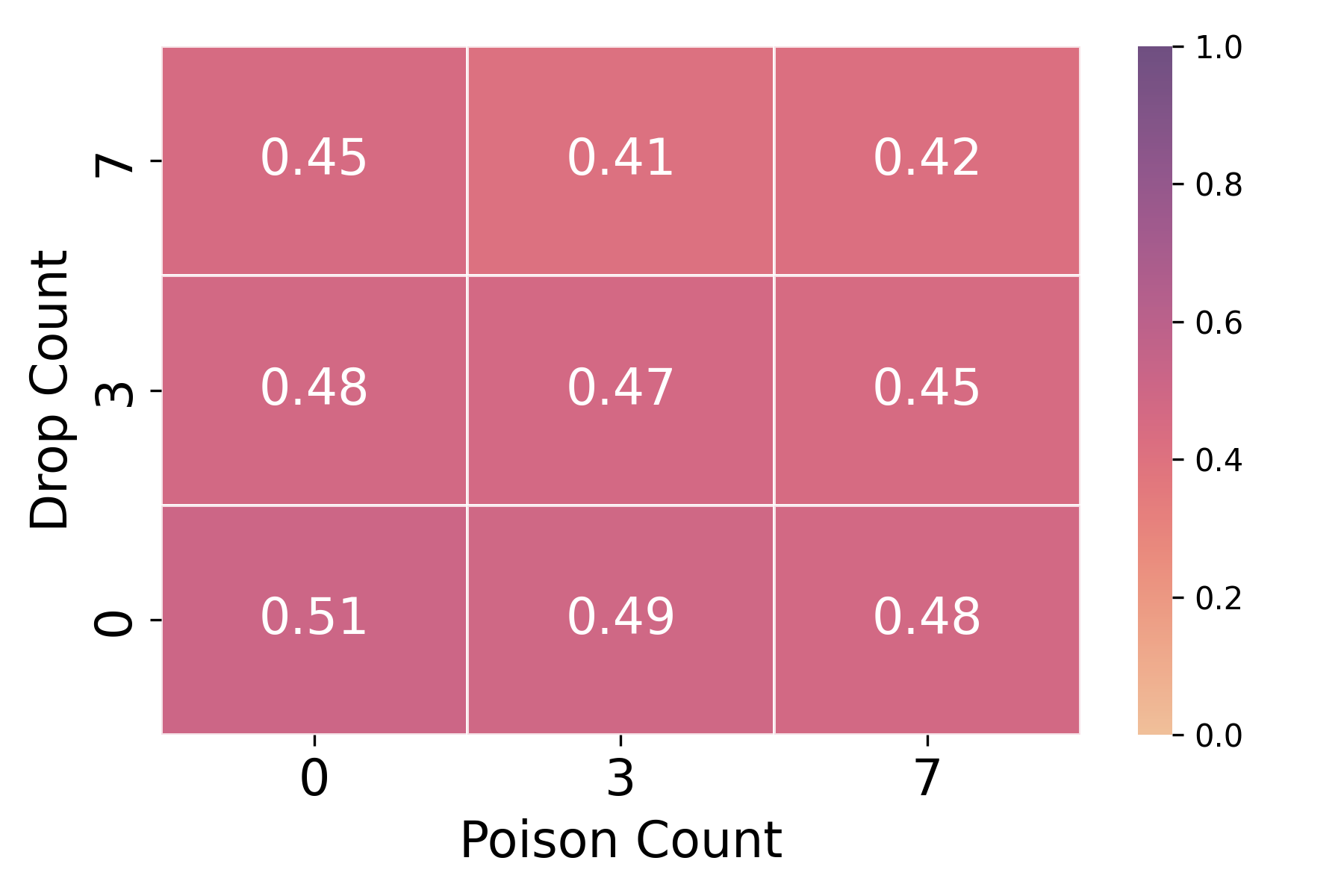}}
    
    \caption{Accuracy on  target class 0 on EMNIST, for $k=15$, $T = 50$, and varying number of dropped and poisoned clients under 3 scenarios: perfect knowledge, \COne, and \CTwo. Left results are without clipping, and right results use a clipping norm of 1. When no clipping is used, model poisoning attack is devastating at small number of poisoned clients. With clipping, model poisoning. Results are similar with those in Figure~\ref{fig:pois_and_drop_clip}.}
    \label{fig:pois_and_drop_clip_50}
\end{figure}

By repeating the experiments with $T=50$, we can observe that it appears that targeted dropping is more effective than poisoning at low round counts, as shown in Fig.~\ref{fig:pois_and_drop_clip_50}
When the model is in its early phase of learning, it needs to see some examples of the target class to begin learning it, and dropping significantly delays the arrival of these examples. Later in training, however, the difference between poisoning and dropping decreases. 
\fi

%% file: sections/eval_visibility.tex
\subsection{Impact of Adversarial Visibility}

\begin{figure}
    \centering
    
    \subfloat[Drop\label{fig:fashion_visibility_drop}]{%
       \includegraphics[width=0.5\linewidth]{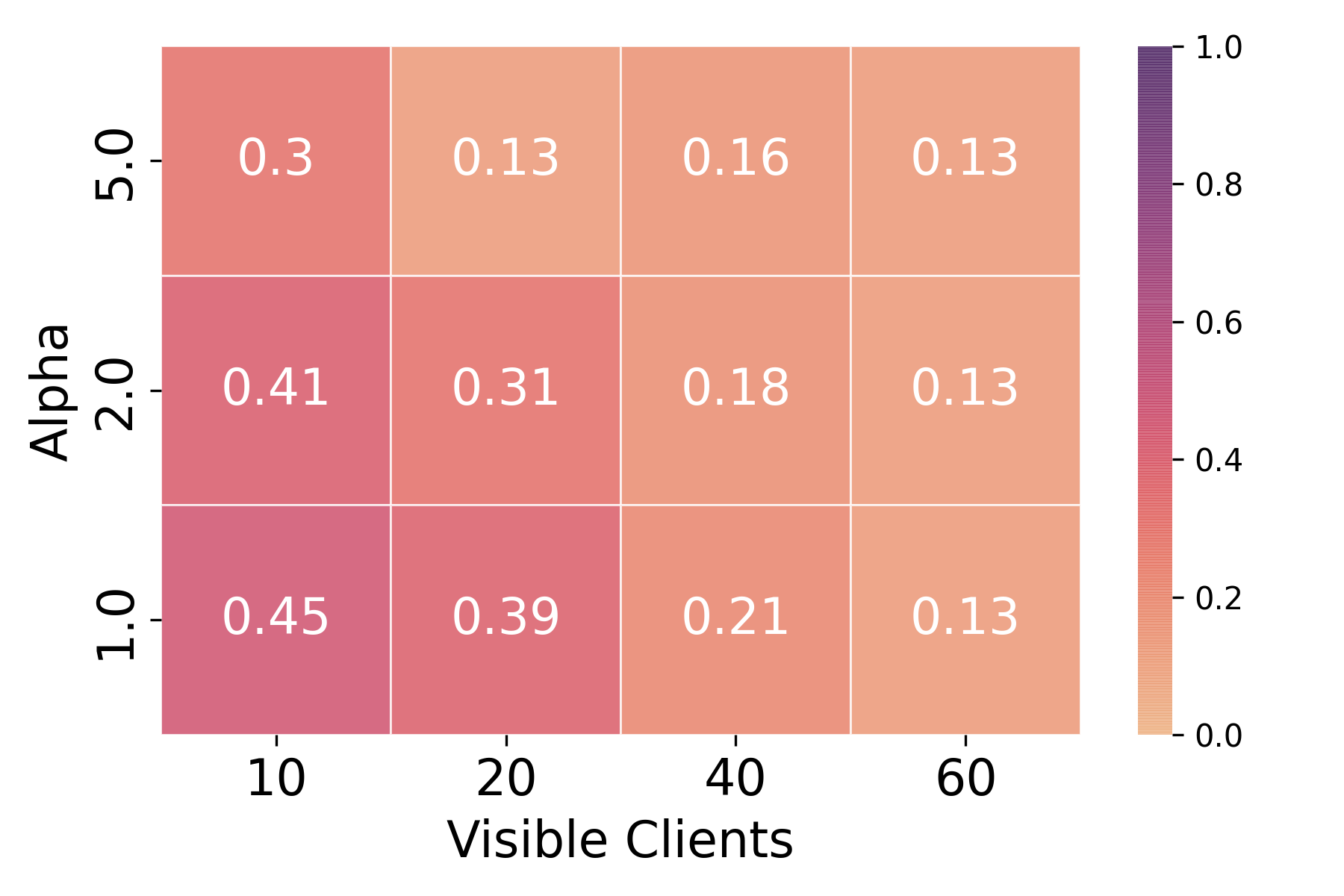}}
    \subfloat[Drop + Poison\label{fig:fashion_visibility_poison}]{%
        \includegraphics[width=0.5\linewidth]{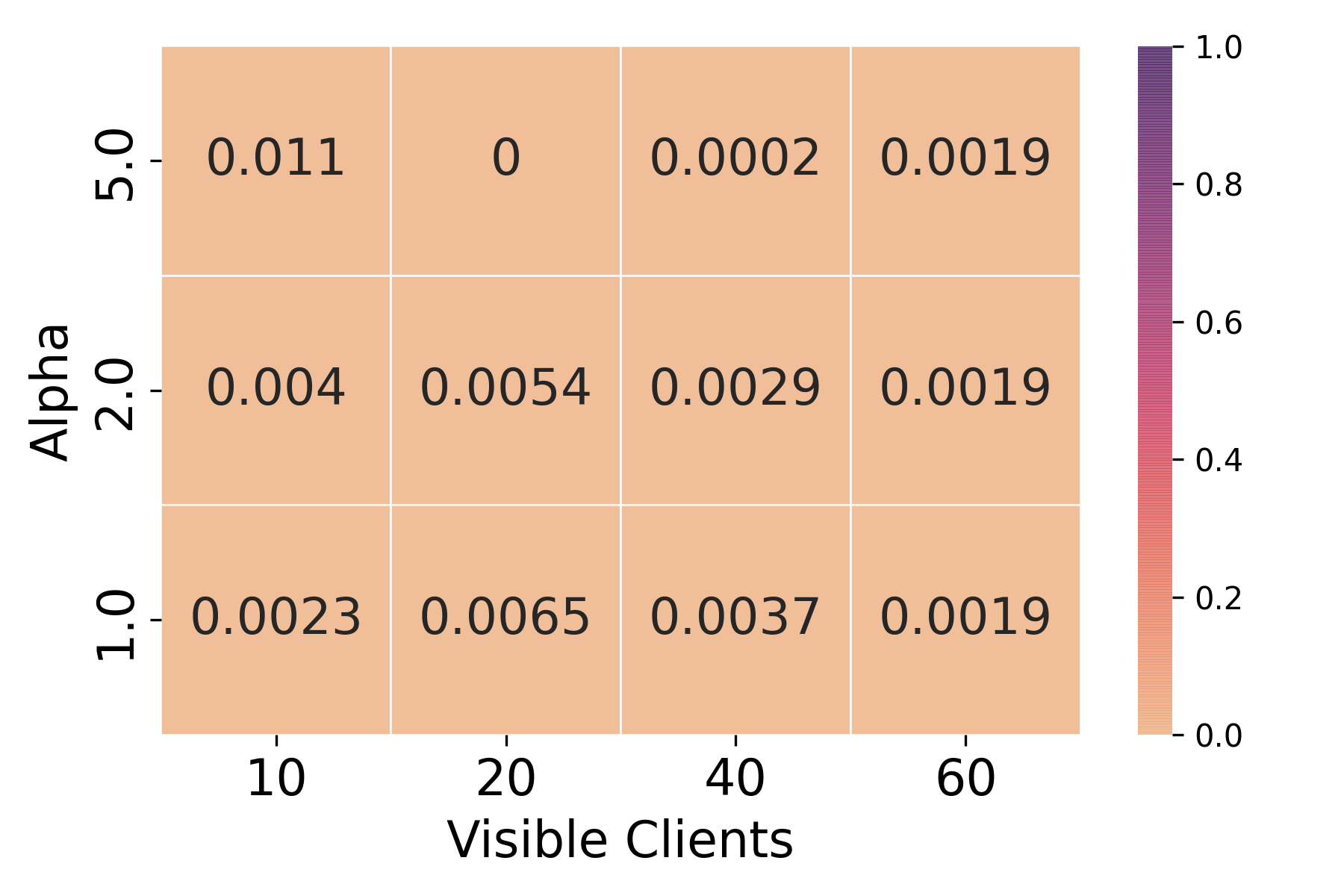}}
    
    \caption{\CVisible. Accuracy on target class 0 on FashionMNIST, for $k=15$, $T = 200$, $k_N = 10$, $k_P = 10$, varying visibility and $\alpha_V$.}
    \label{fig:fashion_visibility}
\vskip -0.18in
\end{figure}

We model a targeted, resource-limited, adversary, \CVisible, by restricting their ability to observe the clients participating to the protocol to a fixed-size subset of the clients, chosen before the attack starts, sampled from a Dirichlet distribution with concentration parameter $\alpha_V$.
$\alpha_V$ controls how likely it is for the sampled visible set to include clients containing data of the target population, with larger values of $\alpha_V$ leading to higher likelihood.
We run the attack on FashionMNIST with a similar setup as in Table~\ref{tab:identification_drop} under the \CTwo\ conditions.
The results reported in Figure~\ref{fig:fashion_visibility_drop} show that, as expected, higher visibility fractions, and larger $\alpha_V$ lead to smaller target accuracy values.
The network adversary is still quite effective even when observing a small fraction of the clients, for instance achieving a $\approx 43.6\%$ relative target accuracy decrease when observing 20 out of 60 clients with $\alpha_V=2$. 
Similarly, Figure~\ref{fig:fashion_visibility_poison} shows that adding poisoning always leads to better accuracy degradation.

%% file: sections/eval_defense.tex
\if\extendedVersion1
\begin{table*}
    \centering
    
    \caption{Accuracy on target class presented at rounds $T / 2$ and $T$, under \COne{} setting. $T = 100$ for EMNIST, $T = 300$ for  FashionMNIST and DBPedia. We consider both Targeted Dropping and Dropping + Poisoning scenarios.}  
    
    \begin{tabular}{cc|c|cc|cccc}
        \multirow{2}{*}{$\boldsymbol{k}$} & \textbf{Target} & \multicolumn{1}{c}{\textbf{No Attack}} & \multicolumn{2}{c}{\textbf{Targeted Drop}} & \multicolumn{4}{c}{\textbf{Targeted Drop + Poison}} \\
        ~ & \textbf{Class} & \textbf{FedAvg} & \textbf{FedAvg} & \textbf{UpSample} & \textbf{FedAvg} & \textbf{Clip} & \textbf{UpSample} & \textbf{Clip + UpSample}  \\
        \hline
        \multicolumn{9}{c}{\multirow{2}{*}{\textbf{EMNIST}}}\\
        \\
        \hline
         \multirow{3}{*}{9}
            & 0 & 0.47/ 0.66  &  0.09/ 0.18 &   0.18/ 0.44 &       0.00/ 0.00 &     0.00/ 0.06 &               0.01/ 0.00 &             0.03/ 0.31 \\
            & 1 &  0.75/ 0.92 &  0.09/ 0.25 &   0.39/ 0.76 &       0.00/ 0.00 &     0.00/ 0.04 &               0.01/ 0.00 &             0.04/ 0.30 \\
            & 9 &  0.43/ 0.56 &  0.01/ 0.17 &   0.09/ 0.40 &       0.00/ 0.00 &     0.00/ 0.01 &               0.01/ 0.00 &             0.00/ 0.23 \\
        \hline
        \multirow{3}{*}{12}
           & 0 & 0.58/ 0.78  &  0.19/ 0.33 &   0.40/ 0.66 &       0.00/ 0.00 &     0.04/ 0.09 &               0.00/ 0.01 &             0.12/ 0.40 \\
           & 1 & 0.86/ 0.95 &   0.25/ 0.55 &   0.49/ 0.85 &       0.00/ 0.00 &     0.01/ 0.04 &               0.04/ 0.01 &             0.17/ 0.60 \\
           & 9 & 0.53/ 0.67 &   0.06/ 0.31 &   0.28/ 0.52 &       0.00/ 0.00 &     0.00/ 0.02 &               0.02/ 0.01 &             0.07/ 0.29 \\
        \hline
        \multirow{3}{*}{15} 
           & 0 & 0.65/ 0.80  &   0.26/ 0.50 &   0.47/ 0.71 &       0.00/ 0.00 &     0.04/ 0.06 &               0.00/ 0.03 &             0.22/ 0.35 \\
           & 1 & 0.91/ 0.96 &   0.47/ 0.79 &   0.83/ 0.94 &       0.00/ 0.00 &     0.06/ 0.18 &               0.05/ 0.04 &             0.53/ 0.51 \\
           & 9 & 0.65/ 0.75 &   0.15/ 0.39 &   0.39/ 0.59 &       0.00/ 0.00 &     0.02/ 0.03 &               0.02/ 0.07 &             0.19/ 0.40 \\
        \hline
        \multicolumn{9}{c}{\multirow{2}{*}{\textbf{FashionMNIST}}}\\
        \\
        \hline
        \multirow{3}{*}{15} 
   & 0 &   0.44/ 0.47 & 0.30/ 0.36 &   0.60/ 0.62 &       0.09/ 0.03 &     0.29/ 0.38 &               0.13/ 0.19 &             0.58/ 0.58 \\
   & 1 &   0.93/ 0.95 & 0.89/ 0.93 &   0.95/ 0.96 &       0.12/ 0.02 &     0.77/ 0.77 &               0.16/ 0.28 &             0.92/ 0.93 \\
   & 9 &   0.88/ 0.87 & 0.81/ 0.81 &   0.92/ 0.93 &       0.13/ 0.04 &     0.70/ 0.62 &               0.09/ 0.19 &             0.90/ 0.89 \\
        \hline
        \multicolumn{9}{c}{\multirow{2}{*}{\textbf{DBPedia}}}\\
        \\
        \hline
        \multirow{3}{*}{15} 
& 0 &   0.45/ 0.54 & 0.11/ 0.10 &   0.75/ 0.82 &       0.00/ 0.00 &     0.00/ 0.00 &               0.02/ 0.01 &             0.65/ 0.77 \\
& 1 &   0.77/ 0.87 & 0.37/ 0.17 &   0.93/ 0.96 &       0.00/ 0.00 &     0.08/ 0.00 &               0.02/ 0.03 &             0.89/ 0.94 \\
& 9 &   0.52/ 0.60 & 0.24/ 0.17 &   0.84/ 0.91 &       0.00/ 0.00 &     0.00/ 0.00 &               0.02/ 0.02 &             0.64/ 0.62 \\
        \hline
    \end{tabular}
    \label{tab:upsample_each}
\end{table*}

\begin{table*}
    \centering
    
    \caption{Accuracy on full test set presented at rounds $T / 2$ and $T$, under \COne{} setting. $T = 100$ for EMNIST, $T = 300$ for  FashionMNIST and DBPedia. We consider both Targeted Dropping and Dropping + Poisoning scenarios.}  
    
    \begin{tabular}{cc|c|cc|cccc}
        \multirow{2}{*}{$\boldsymbol{k}$} & \textbf{Target} & \multicolumn{1}{c}{\textbf{No Attack}} & \multicolumn{2}{c}{\textbf{Targeted Drop}} & \multicolumn{4}{c}{\textbf{Targeted Drop + Poison}} \\
        ~ & \textbf{Class} & \textbf{FedAvg} & \textbf{FedAvg} & \textbf{UpSample} & \textbf{FedAvg} & \textbf{Clip} & \textbf{UpSample} & \textbf{Clip + UpSample}  \\
        \hline
        \multicolumn{9}{c}{\multirow{2}{*}{\textbf{EMNIST}}}\\
        \\
        \hline
        \multirow{3}{*}{9} 
           & 0 &  0.92/ 0.95 &  0.88/ 0.90 &   0.89/ 0.93 &       0.86/ 0.87 &     0.85/ 0.88 &               0.86/ 0.87 &             0.86/ 0.91 \\
           & 1 &  0.94/ 0.96 &  0.87/ 0.90 &   0.90/ 0.95 &       0.85/ 0.85 &     0.84/ 0.86 &               0.85/ 0.86 &             0.85/ 0.89 \\
           & 9 &  0.91/ 0.93 &  0.87/ 0.89 &   0.88/ 0.92 &       0.86/ 0.87 &     0.86/ 0.87 &               0.86/ 0.87 &             0.86/ 0.90 \\
        \hline
        \multirow{3}{*}{12} 
           & 0 &  0.93/ 0.96 &  0.89/ 0.92 &   0.91/ 0.95 &       0.86/ 0.87 &     0.86/ 0.89 &               0.86/ 0.87 &             0.87/ 0.92 \\
           & 1 &  0.95/ 0.97 &  0.89/ 0.93 &   0.91/ 0.96 &       0.85/ 0.86 &     0.84/ 0.86 &               0.85/ 0.86 &             0.86/ 0.92 \\
           & 9 &  0.92/ 0.95 &  0.87/ 0.91 &   0.90/ 0.93 &       0.86/ 0.87 &     0.86/ 0.87 &               0.68/ 0.68 &             0.87/ 0.90 \\
        \hline
        \multirow{3}{*}{15} 
           & 0 &  0.94/ 0.96 &  0.90/ 0.93 &   0.92/ 0.95 &       0.86/ 0.87 &     0.86/ 0.88 &               0.86/ 0.88 &             0.89/ 0.91 \\
           & 1 &  0.95/ 0.97 &  0.91/ 0.95 &   0.94/ 0.96 &       0.85/ 0.86 &     0.85/ 0.88 &               0.85/ 0.86 &             0.90/ 0.91 \\
           & 9 &  0.94/ 0.95 &  0.88/ 0.92 &   0.91/ 0.94 &       0.86/ 0.87 &     0.86/ 0.87 &               0.87/ 0.88 &             0.88/ 0.92 \\
        \hline
        \multicolumn{9}{c}{\multirow{2}{*}{\textbf{FashionMNIST}}}\\
        \\
        \hline
        \multirow{3}{*}{15} 
         & 0 & 0.81/ 0.83 &  0.79/ 0.82 &   0.81/ 0.84 &       0.76/ 0.78 &     0.79/ 0.82 &               0.76/ 0.79 &             0.81/ 0.84 \\
         & 1 & 0.83/ 0.85 &  0.82/ 0.85 &   0.82/ 0.85 &       0.74/ 0.76 &     0.81/ 0.83 &               0.75/ 0.78 &             0.82/ 0.84 \\
         & 9 & 0.82/ 0.85 &  0.82/ 0.84 &   0.83/ 0.85 &       0.74/ 0.76 &     0.80/ 0.82 &               0.74/ 0.77 &             0.82/ 0.84 \\
        \hline
        \multicolumn{9}{c}{\multirow{2}{*}{\textbf{DBPedia}}}\\
        \\
        \hline
        \multirow{3}{*}{15} 
        & 0 & 0.93/ 0.94 &  0.91/ 0.91 &   0.95/ 0.96 &       0.89/ 0.90 &     0.90/ 0.90 &               0.89/ 0.90 &             0.94/ 0.96 \\
        & 1 & 0.95/ 0.96 &  0.92/ 0.91 &   0.96/ 0.97 &       0.88/ 0.89 &     0.90/ 0.90 &               0.89/ 0.89 &             0.95/ 0.96 \\
        & 9 & 0.93/ 0.94 &  0.91/ 0.91 &   0.95/ 0.96 &       0.89/ 0.90 &     0.89/ 0.90 &               0.89/ 0.90 &             0.93/ 0.94 \\
        \hline
    \end{tabular}
    \label{tab:upsample_global_each}
\end{table*}

\begin{table*}
    \centering
    
    \caption{Accuracy on target class presented at rounds $T / 2$ and $T$, under \CTwo{} setting. $T = 100$ for EMNIST, $T = 300$ for  FashionMNIST and DBPedia. We consider both Targeted Dropping and Dropping + Poisoning scenarios.}  
    
    \begin{tabular}{cc|c|cc|cccc}
        \multirow{2}{*}{$\boldsymbol{k}$} & \textbf{Target} & \multicolumn{1}{c}{\textbf{No Attack}} & \multicolumn{2}{c}{\textbf{Targeted Drop}} & \multicolumn{4}{c}{\textbf{Targeted Drop + Poison}} \\
        ~ & \textbf{Class} & \textbf{FedAvg} & \textbf{FedAvg} & \textbf{UpSample} & \textbf{FedAvg} & \textbf{Clip} & \textbf{UpSample} & \textbf{Clip + UpSample}  \\
        \hline
        \multicolumn{9}{c}{\multirow{2}{*}{\textbf{EMNIST}}}\\
        \\
        \hline
        \multirow{3}{*}{9} & 0 &0.47/ 0.66 &    0.32/ 0.52 &   0.59/ 0.76 &       0.01/ 0.00 &     0.14/ 0.40 &               0.10/ 0.04 &             0.46/ 0.68 \\
           & 1 &0.75/ 0.92 &    0.58/ 0.76 &   0.88/ 0.96 &       0.04/ 0.00 &     0.24/ 0.52 &               0.16/ 0.04 &             0.67/ 0.88 \\
           & 9 &0.43/ 0.56 &    0.31/ 0.46 &   0.54/ 0.70 &       0.01/ 0.00 &     0.03/ 0.25 &               0.06/ 0.05 &             0.37/ 0.57 \\
        \hline
        \multirow{3}{*}{12} & 0 &0.58/ 0.78 &    0.48/ 0.69 &   0.75/ 0.85 &       0.00/ 0.00 &     0.29/ 0.36 &               0.02/ 0.06 &             0.60/ 0.77 \\
           & 1 & 0.86/ 0.95 &    0.77/ 0.91 &   0.94/ 0.97 &       0.00/ 0.00 &     0.47/ 0.43 &               0.20/ 0.12 &             0.82/ 0.92 \\
           & 9 & 0.53/ 0.67 &   0.41/ 0.56 &   0.68/ 0.78 &       0.00/ 0.01 &     0.18/ 0.27 &               0.15/ 0.07 &             0.50/ 0.67 \\
        \hline
        \multirow{3}{*}{15} & 0 &0.65/ 0.80 &    0.60/ 0.76 &   0.81/ 0.89 &       0.01/ 0.00 &     0.37/ 0.44 &               0.01/ 0.05 &             0.62/ 0.71 \\
           & 1 & 0.91/ 0.96 &   0.88/ 0.94 &   0.96/ 0.97 &       0.04/ 0.00 &     0.60/ 0.48 &               0.12/ 0.06 &             0.90/ 0.94 \\
           & 9 & 0.65/ 0.75 &   0.50/ 0.58 &   0.76/ 0.85 &       0.03/ 0.01 &     0.30/ 0.35 &               0.10/ 0.07 &             0.53/ 0.71 \\
        \hline
        \multicolumn{9}{c}{\multirow{2}{*}{\textbf{FashionMNIST}}}\\
        \\
        \hline
        \multirow{3}{*}{15} 
& 0 &  0.44/ 0.47 & 0.40/ 0.39 &   0.69/ 0.71 &       0.13/ 0.02 &     0.38/ 0.36 &               0.25/ 0.17 &             0.64/ 0.69 \\
& 1 &  0.93/ 0.95 & 0.92/ 0.94 &   0.95/ 0.96 &       0.14/ 0.03 &     0.86/ 0.81 &               0.26/ 0.24 &             0.94/ 0.95 \\
& 9 &  0.88/ 0.87 & 0.85/ 0.82 &   0.93/ 0.94 &       0.20/ 0.05 &     0.77/ 0.59 &               0.19/ 0.23 &             0.91/ 0.92 \\
        \hline
        \multicolumn{9}{c}{\multirow{2}{*}{\textbf{DBPedia}}}\\
        \\
        \hline
        \multirow{3}{*}{15} 
& 0 & 0.45/ 0.54 &  0.16/ 0.12 &   0.79/ 0.84 &       0.00/ 0.00 &     0.00/ 0.00 &               0.03/ 0.01 &             0.73/ 0.79 \\
& 1 & 0.77/ 0.87 &  0.39/ 0.22 &   0.95/ 0.97 &       0.00/ 0.00 &     0.07/ 0.00 &               0.10/ 0.05 &             0.92/ 0.95 \\
& 9 & 0.52/ 0.60 &  0.31/ 0.12 &   0.88/ 0.94 &       0.00/ 0.00 &     0.00/ 0.00 &               0.02/ 0.01 &             0.75/ 0.76 \\
        \hline
    \end{tabular}

    \label{tab:upsample_mean}
\end{table*}

\begin{table*}
    \centering
    
    \caption{Accuracy on full test set presented at rounds $T / 2$ and $T$, under \CTwo{} setting. $T = 100$ for EMNIST, $T = 300$ for  FashionMNIST and DBPedia. We consider both Targeted Dropping and Dropping + Poisoning scenarios.}  
    
    \begin{tabular}{cc|c|cc|cccc}
        \multirow{2}{*}{$\boldsymbol{k}$} & \textbf{Target} & \multicolumn{1}{c}{\textbf{No Attack}} & \multicolumn{2}{c}{\textbf{Targeted Drop}} & \multicolumn{4}{c}{\textbf{Targeted Drop + Poison}} \\
        ~ & \textbf{Class} & \textbf{FedAvg} & \textbf{FedAvg} & \textbf{UpSample} & \textbf{FedAvg} & \textbf{Clip} & \textbf{UpSample} & \textbf{Clip + UpSample}  \\
        \hline
        \multicolumn{9}{c}{\multirow{2}{*}{\textbf{EMNIST}}}\\
        \\
        \hline
        \multirow{3}{*}{9} 
           & 0 &  0.92/ 0.95 & 0.91/ 0.94 &   0.93/ 0.95 &       0.86/ 0.87 &     0.88/ 0.92 &               0.87/ 0.87 &             0.91/ 0.94 \\
           & 1 &  0.94/ 0.96 & 0.92/ 0.95 &   0.95/ 0.97 &       0.85/ 0.86 &     0.88/ 0.91 &               0.86/ 0.86 &             0.92/ 0.95 \\
           & 9 &  0.91/ 0.93 & 0.90/ 0.92 &   0.93/ 0.95 &       0.86/ 0.87 &     0.86/ 0.90 &               0.87/ 0.88 &             0.90/ 0.93 \\
        \hline
        \multirow{3}{*}{12} 
           & 0 &  0.93/ 0.96 & 0.92/ 0.95 &   0.94/ 0.96 &       0.86/ 0.87 &     0.90/ 0.92 &               0.87/ 0.88 &             0.92/ 0.95 \\
           & 1 &  0.95/ 0.97 & 0.94/ 0.96 &   0.95/ 0.97 &       0.85/ 0.86 &     0.90/ 0.90 &               0.87/ 0.87 &             0.93/ 0.95 \\
           & 9 &  0.92/ 0.95 & 0.91/ 0.94 &   0.94/ 0.96 &       0.86/ 0.87 &     0.88/ 0.90 &               0.88/ 0.88 &             0.91/ 0.94 \\
        \hline
        \multirow{3}{*}{15} 
           & 0 &  0.94/ 0.96 &  0.93/ 0.95 &   0.95/ 0.96 &       0.86/ 0.87 &     0.90/ 0.92 &               0.87/ 0.88 &             0.92/ 0.95 \\
           & 1 &  0.95/ 0.97 &  0.95/ 0.96 &   0.95/ 0.97 &       0.85/ 0.86 &     0.91/ 0.90 &               0.86/ 0.86 &             0.94/ 0.96 \\
           & 9 &  0.94/ 0.95 &  0.92/ 0.94 &   0.94/ 0.96 &       0.87/ 0.87 &     0.89/ 0.91 &               0.88/ 0.88 &             0.92/ 0.95 \\
        \hline
        \multicolumn{9}{c}{\multirow{2}{*}{\textbf{FashionMNIST}}}\\
        \\
        \hline
        \multirow{3}{*}{15} 
   & 0 &  0.81/ 0.83 & 0.80/ 0.83 &   0.82/ 0.85 &       0.76/ 0.78 &     0.80/ 0.82 &               0.77/ 0.78 &             0.81/ 0.84 \\
  & 1 &  0.83/ 0.85 & 0.83/ 0.85 &   0.82/ 0.85 &       0.75/ 0.76 &     0.82/ 0.83 &               0.76/ 0.78 &             0.82/ 0.85 \\
  & 9 &  0.82/ 0.85 & 0.82/ 0.84 &   0.83/ 0.85 &       0.74/ 0.76 &     0.80/ 0.82 &               0.75/ 0.78 &             0.82/ 0.85 \\
        \hline
        \multicolumn{9}{c}{\multirow{2}{*}{\textbf{DBPedia}}}\\
        \\
        \hline
        \multirow{3}{*}{15} 
& 0 &  0.93/ 0.94 & 0.91/ 0.91 &   0.95/ 0.96 &       0.89/ 0.90 &     0.90/ 0.90 &               0.90/ 0.90 &             0.95/ 0.96 \\
& 1 &  0.95/ 0.96  & 0.92/ 0.92 &   0.96/ 0.97 &       0.88/ 0.89 &     0.89/ 0.90 &               0.89/ 0.89 &             0.95/ 0.97 \\
& 9 &  0.93/ 0.94 & 0.92/ 0.91 &   0.95/ 0.96 &       0.89/ 0.90 &     0.89/ 0.90 &               0.89/ 0.89 &             0.93/ 0.95 \\
        \hline
    \end{tabular}

    \label{tab:upsample_global_mean}
\end{table*}

\fi

\subsection{Defense Evaluation}
\label{sec:eval_defense}    
 
\if\extendedVersion1   
In Tables~\ref{tab:upsample_each}, ~\ref{tab:upsample_mean}, 
\else
In Table~\ref{tab:upsample_combined},
\fi
we evaluate the defense strategies under \COne\ and \CTwo, using targeted dropping and poisoning attacks.
We set the number of dropped ($k_N$) and poisoned clients ($k_P$) to $ 2k / 3$ for EMNIST, and $k / 3$ for FashionMNIST and DBPedia. These parameters result in strong attacks, as the target accuracy is below 4\% under targeted dropping and poisoning, when no mitigation is used. 
UpSampling is successful in mitigating targeted dropping, and, when combined with Clipping, achieves high accuracy against the powerful combined attack. For instance, on EMNIST's class 1 with $k=9$, the update dropping attack causes accuracy to decrease from 92\% to 25\% under \COne, and UpSampling manages to restore the target accuracy to 76\%. 

The UpSampling defense is very effective under the \CTwo\ setting, due to knowledge asymmetry, i.e. the attacker receives aggregated updates, while the defender observes individual updates. Thus, UpSampling, defending against the dropping attack, improves the target model accuracy by an average (over the 3 classes) of 6.33\% on EMNIST, 10.66\% on FashionMNIST, and 24.66\% on DBPedia for $k=15$ compared to the original accuracy.
Even under both targeted dropping and model poisoning attacks, the combined UpSampling defense and Clipping results in an average decrease of 5\% on EMNIST and average increase of 9\% on FashionMNIST and 16.66\% on DBPedia over the 3 classes compared to the original accuracy. Interestingly, classes with lower original accuracy benefit more from the UpSampling strategy, with improvements as high as 34\% (on DBPedia for class 9, original accuracy is increased from 60\% to 94\% with UpSampling under targeted dropping).


We observed that under-represented classes (in terms of number of clients holding samples from those classes) are impacted more by our attacks. To alleviate this problem, the server could identify new clients with data from the populations of interest, and add them to the set of clients participating in the FL protocol. In cross-device FL settings, servers typically have access to a large number of clients, and can make decisions on expanding the set of participating clients to improve accuracy on under-represented populations.
\if\extendedVersion1

\myparagraph{Privacy-preserving FL} So far, we have discussed settings in which the server receives local model updates in all rounds of the FL protocol. 
However, to protect client privacy, it is common to deploy privacy-preserving FL protocols, based on Multi-Party Computation (MPC), such as~\cite{bonawitzPracticalSecureAggregation2017, fereidooniSAFELearnSecureAggregation2021}. 
In MPC implementations, multiple parties will be involved in aggregation and the server only receives the global model at the end of each iteration. 
The server has therefore the same knowledge as the network-level adversary under encrypted communication when running the client identification protocol from Algorithm~\ref{alg:client_id}.

In Table~\ref{tab:upsample_c3}, we present attack and defense results for this challenging setting.
Similarly to the previous tables, we set the parameters to $k_N = k_P = 2k / 3$ for EMNIST, and $k_N = k_P = k / 3$ for both FashionMNIST and DBPedia.
While under this setting there is essentially no difference in the effectiveness of the attacker models we consider, the defensive performance varies due to the limited knowledge.
The results we observe for the UpSample and Clip + UpSample defensive mechanisms are, as we would expect, somewhere in between the \CTwo{} and \COne{} scenarios.
For instance for EMNIST, with $k=15$ on class 1 we obtain an average accuracy level of 0.84, considerably higher than the 0.51 obtained under \COne{}, but also not as high as the average of 0.94 obtained with \CTwo{}, which is essentially the same result obtained without any attack.


\begin{table*}
    \centering
    
    \caption{Accuracy on target class presented at rounds $T / 2$ and $T$, under the \CThree{} scenario. $T = 100$ for EMNIST, $T = 300$ for  FashionMNIST and DBPedia. We consider both Targeted Dropping and Dropping + Poisoning scenarios.}  
    
    \begin{tabular}{cc|c|cc|cccc}
        \multirow{2}{*}{$\boldsymbol{k}$} & \textbf{Target} & \multicolumn{1}{c}{\textbf{No Attack}} & \multicolumn{2}{c}{\textbf{Targeted Drop}} & \multicolumn{4}{c}{\textbf{Targeted Drop + Poison}} \\
        ~ & \textbf{Class} & \textbf{FedAvg} & \textbf{FedAvg} & \textbf{UpSample} & \textbf{FedAvg} & \textbf{Clip} & \textbf{UpSample} & \textbf{Clip + UpSample}  \\
        \hline
        \multicolumn{9}{c}{\multirow{2}{*}{\textbf{EMNIST}}}\\
        \\
        \hline
        \multirow{3}{*}{9} & 0 &0.47/ 0.66&   0.32/ 0.52 &   0.39/ 0.67 &       0.01/ 0.00 &     0.14/ 0.40 &               0.02/ 0.00 &             0.25/ 0.55 \\
           & 1 &0.75/ 0.92&    0.58/ 0.76 &   0.72/ 0.93 &       0.04/ 0.00 &     0.24/ 0.52 &               0.01/ 0.02 &             0.49/ 0.73 \\
           & 9 &0.43/ 0.56&    0.31/ 0.46 &   0.38/ 0.60 &       0.01/ 0.00 &     0.03/ 0.25 &               0.03/ 0.00 &             0.14/ 0.50 \\
        \hline
        \multirow{3}{*}{12} & 0 &0.58/ 0.78&    0.48/ 0.69 &   0.50/ 0.75 &       0.00/ 0.00 &     0.29/ 0.36 &               0.02/ 0.02 &             0.40/ 0.62 \\
           & 1 &0.86/ 0.95&    0.77/ 0.91 &   0.85/ 0.96 &       0.00/ 0.00 &     0.47/ 0.43 &               0.00/ 0.03 &             0.63/ 0.78 \\
           & 9 &0.53/ 0.67&    0.41/ 0.56 &   0.54/ 0.72 &       0.00/ 0.01 &     0.18/ 0.27 &               0.02/ 0.00 &             0.27/ 0.51 \\
        \hline
        \multirow{3}{*}{15} & 0 &0.65/ 0.80&    0.60/ 0.76 &   0.69/ 0.84 &       0.01/ 0.00 &     0.37/ 0.44 &               0.01/ 0.04 &             0.43/ 0.63 \\
           & 1 &0.91/ 0.96&    0.87/ 0.93 &   0.92/ 0.96 &       0.04/ 0.00 &     0.60/ 0.48 &               0.01/ 0.00 &             0.73/ 0.84 \\
           & 9 &0.64/ 0.75&    0.50/ 0.58 &   0.59/ 0.80 &       0.03/ 0.01 &     0.30/ 0.35 &               0.02/ 0.00 &             0.38/ 0.61 \\
    \hline
        \multicolumn{9}{c}{\multirow{2}{*}{\textbf{FashionMNIST}}}\\ 
        \\
        \hline
        \multirow{3}{*}{15} 
& 0 &  0.44/ 0.46 &  0.40/ 0.41 &   0.61/ 0.70 &       0.12/ 0.04 &     0.39/ 0.35 &               0.04/ 0.05 &             0.60/ 0.65 \\
   & 1 & 0.93/ 0.95 &  0.92/ 0.94 &   0.95/ 0.96 &       0.11/ 0.03 &     0.86/ 0.81 &               0.14/ 0.06 &             0.93/ 0.95 \\
   & 9 & 0.88/ 0.87 &  0.85/ 0.82 &   0.93/ 0.93 &       0.19/ 0.05 &     0.75/ 0.60 &               0.08/ 0.04 &             0.86/ 0.85 \\
        \hline
        \multicolumn{9}{c}{\multirow{2}{*}{\textbf{DBPedia}}}\\
        \\
        \hline
        \multirow{3}{*}{15} 
            & 0 &  0.45/ 0.54 & 0.14/ 0.12 &   0.70/ 0.81 &       0.00/ 0.00 &     0.00/ 0.00 &               0.00/ 0.00 &             0.61/ 0.77 \\
            & 1 &  0.77/ 0.87 & 0.48/ 0.32 &   0.92/ 0.96 &       0.00/ 0.00 &     0.07/ 0.00 &               0.00/ 0.03 &             0.87/ 0.93 \\
            & 9 &  0.52/ 0.60 & 0.30/ 0.13 &   0.82/ 0.91 &       0.00/ 0.00 &     0.00/ 0.00 &               0.00/ 0.00 &             0.53/ 0.70 \\
        
        \hline
    \end{tabular}
    \label{tab:upsample_c3}
\end{table*}

\begin{table*}
    \centering
    
    \caption{Accuracy on full test set presented at rounds $T / 2$ and $T$, under the \CThree{} scenario. $T = 100$ for EMNIST, $T = 300$ for  FashionMNIST and DBPedia. We consider both Targeted Dropping and Dropping + Poisoning scenarios.}  
    
    \begin{tabular}{cc|c|cc|cccc}
        \multirow{2}{*}{$\boldsymbol{k}$} & \textbf{Target} & \multicolumn{1}{c}{\textbf{No Attack}} & \multicolumn{2}{c}{\textbf{Targeted Drop}} & \multicolumn{4}{c}{\textbf{Targeted Drop + Poison}} \\
        ~ & \textbf{Class} & \textbf{FedAvg} & \textbf{FedAvg} & \textbf{UpSample} & \textbf{FedAvg} & \textbf{Clip} & \textbf{UpSample} & \textbf{Clip + UpSample}  \\
        \hline
        \multicolumn{9}{c}{\multirow{2}{*}{\textbf{EMNIST}}}\\
        \\
        \hline
        \multirow{3}{*}{9} 
           & 0 & 0.92/ 0.95 &  0.91/ 0.94 &   0.91/ 0.95 &       0.86/ 0.87 &     0.88/ 0.92 &               0.87/ 0.87 &             0.89/ 0.93 \\
           & 1 & 0.94/ 0.96 &  0.92/ 0.95 &   0.93/ 0.96 &       0.85/ 0.86 &     0.88/ 0.91 &               0.85/ 0.86 &             0.90/ 0.93 \\
           & 9 & 0.91/ 0.93 &  0.90/ 0.92 &   0.91/ 0.94 &       0.86/ 0.87 &     0.86/ 0.90 &               0.87/ 0.87 &             0.88/ 0.93 \\
        \hline
        \multirow{3}{*}{12} 
           & 0 &  0.93/ 0.96 &  0.92/ 0.95 &   0.92/ 0.95 &       0.86/ 0.87 &     0.90/ 0.92 &               0.86/ 0.87 &             0.91/ 0.94 \\
           & 1 &  0.95/ 0.97 &  0.94/ 0.96 &   0.95/ 0.97 &       0.85/ 0.86 &     0.90/ 0.90 &               0.85/ 0.86 &             0.91/ 0.94 \\
           & 9 &  0.92/ 0.95 &  0.91/ 0.94 &   0.93/ 0.95 &       0.86/ 0.87 &     0.88/ 0.90 &               0.87/ 0.68 &             0.89/ 0.93 \\
        \hline
        \multirow{3}{*}{15} 
           & 0 & 0.94/ 0.96 &  0.93/ 0.95 &   0.94/ 0.96 &       0.86/ 0.87 &     0.90/ 0.92 &               0.87/ 0.88 &             0.91/ 0.94 \\
           & 1 & 0.95/ 0.97 &  0.95/ 0.96 &   0.95/ 0.97 &       0.85/ 0.86 &     0.91/ 0.90 &               0.85/ 0.86 &             0.92/ 0.95 \\
           & 9 & 0.94/ 0.95 &  0.92/ 0.94 &   0.93/ 0.96 &       0.87/ 0.87 &     0.89/ 0.91 &               0.87/ 0.87 &             0.90/ 0.94 \\
    \hline
        \multicolumn{9}{c}{\multirow{2}{*}{\textbf{FashionMNIST}}}\\ 
        \\
        \hline
        \multirow{3}{*}{15} 
        & 0 & 0.81/ 0.83 &  0.80/ 0.83 &   0.82/ 0.85 &       0.76/ 0.78 &     0.80/ 0.82 &               0.75/ 0.77 &             0.81/ 0.84 \\
        & 1 & 0.83/ 0.85 &  0.83/ 0.85 &   0.82/ 0.85 &       0.74/ 0.76 &     0.82/ 0.83 &               0.75/ 0.76 &             0.82/ 0.85 \\
        & 9 & 0.82/ 0.85 &  0.82/ 0.84 &   0.83/ 0.85 &       0.74/ 0.76 &     0.80/ 0.82 &               0.74/ 0.76 &             0.81/ 0.84 \\
        \hline
        \multicolumn{9}{c}{\multirow{2}{*}{\textbf{DBPedia}}}\\
        \\
        \hline
        \multirow{3}{*}{15} 
        & 0 & 0.93/ 0.94 &  0.91/ 0.91 &   0.95/ 0.96 &       0.89/ 0.89 &     0.90/ 0.90 &               0.89/ 0.90 &             0.94/ 0.96 \\
        & 1 & 0.95/ 0.96 &  0.93/ 0.92 &   0.96/ 0.97 &       0.88/ 0.89 &     0.89/ 0.90 &               0.88/ 0.89 &             0.95/ 0.96 \\
        & 9 & 0.93/ 0.94 &  0.92/ 0.91 &   0.95/ 0.96 &       0.89/ 0.90 &     0.89/ 0.90 &               0.89/ 0.89 &             0.92/ 0.95 \\
        
        \hline
    \end{tabular}
    \label{tab:upsample_global_c3}
\end{table*}

\else
We have also evaluated the attacks and defenses against FL protocols using secure aggregation~\cite{bonawitzPracticalSecureAggregation2017}. Here, both the network-level adversary and the server only observe aggregated model updates. Even in this challenging setting, the server can identify clients contributing mostly to the target class, making  UpSampling combined with Clipping quite effective. Results are omitted for space limitation. 


\fi

\if\extendedVersion0
\begin{table*}
    \scriptsize
    \centering
    
    \caption{Target class accuracy, under \COne\ and \CTwo, at $T = 100$ for EMNIST, and $T = 300$ for  FashionMNIST and DBPedia. Targeted Dropping and Targeted Dropping + Poisoning attack scenarios, and both Clipping ({\bf Clip}) and UpSampling ({\bf UpS}) defenses.
    }  
    
    \begin{tabular}{cc | c | cc | cc | cccc | cccc}
        
        \multirow{3}{*}{$\boldsymbol{k}$} & \textbf{Target} & \textbf{No Attack} & \multicolumn{4}{c|}{\textbf{Targeted Drop}} & \multicolumn{8}{c}{\textbf{Targeted Drop + Poison}} \\

        ~ & \textbf{Class} & \textbf{FedAvg} & \multicolumn{2}{c}{\COne} & \multicolumn{2}{c|}{\CTwo} & \multicolumn{4}{c}{\COne} & \multicolumn{4}{c}{\CTwo} \\ 

        ~ & ~ & ~ & \textbf{FedAvg} & \textbf{UpS} & \textbf{FedAvg} & \textbf{UpS} & \textbf{FedAvg} & \textbf{Clip} & \textbf{UpS} & \textbf{Clip + UpS} & \textbf{FedAvg} & \textbf{Clip} & \textbf{UpS} & \textbf{Clip + UpS} \\
        
        \hline
        \multicolumn{15}{c}{\multirow{2}{*}{\textbf{EMNIST}}}\\
        \\
        \hline
         \multirow{3}{*}{9}
            ~ & 0    & 0.66    & 0.18 & 0.44    & 0.52 & 0.76    & 0.00 & 0.06 & 0.00 & 0.31    & 0.00 & 0.40 & 0.04 & 0.68 \\
            ~ & 1    & 0.92    & 0.25 & 0.76    & 0.76 & 0.96    & 0.00 & 0.04 & 0.00 & 0.30    & 0.00 & 0.52 & 0.04 & 0.88 \\
            ~ & 9    & 0.56    & 0.17 & 0.40    & 0.46 & 0.70    & 0.00 & 0.01 & 0.00 & 0.23    & 0.00 & 0.25 & 0.05 & 0.57 \\
        \hline
        \multirow{3}{*}{12}
            ~ & 0    & 0.78    & 0.33 & 0.66    & 0.69 & 0.85    & 0.00 & 0.09 & 0.01 & 0.40    & 0.00 & 0.36 & 0.06 & 0.77 \\
            ~ & 1    & 0.95    & 0.55 & 0.85    & 0.91 & 0.97    & 0.00 & 0.04 & 0.01 & 0.60    & 0.00 & 0.43 & 0.12 & 0.92 \\
            ~ & 9    & 0.67    & 0.31 & 0.52    & 0.56 & 0.78    & 0.00 & 0.02 & 0.01 & 0.29    & 0.01 & 0.27 & 0.07 & 0.67 \\
        \hline
        \multirow{3}{*}{15} 
            ~ & 0    & 0.80    & 0.50 & 0.71    & 0.76 & 0.89    & 0.00 & 0.06 & 0.03 & 0.35    & 0.00 & 0.44 & 0.05 & 0.71 \\
            ~ & 1    & 0.96    & 0.79 & 0.94    & 0.94 & 0.97    & 0.00 & 0.18 & 0.04 & 0.51    & 0.00 & 0.48 & 0.06 & 0.94 \\
            ~ & 9    & 0.75    & 0.39 & 0.59    & 0.58 & 0.85    & 0.00 & 0.03 & 0.07 & 0.40    & 0.01 & 0.35 & 0.07 & 0.71 \\
        \hline

        \multicolumn{15}{c}{\multirow{2}{*}{\textbf{FashionMNIST}}}\\
        \\
        \hline
        \multirow{3}{*}{15} 
            ~ & 0    & 0.47    & 0.36 & 0.62    & 0.39 & 0.71    & 0.03 & 0.38 & 0.19 & 0.58    & 0.02 & 0.36 & 0.17 & 0.69 \\
            ~ & 1    & 0.95    & 0.93 & 0.96    & 0.94 & 0.96    & 0.02 & 0.77 & 0.28 & 0.93    & 0.03 & 0.81 & 0.24 & 0.95 \\
            ~ & 9    & 0.87    & 0.81 & 0.93    & 0.82 & 0.94    & 0.04 & 0.62 & 0.19 & 0.89    & 0.05 & 0.59 & 0.23 & 0.92 \\
        \hline
        
        \multicolumn{15}{c}{\multirow{2}{*}{\textbf{DBPedia}}}\\
        \\
        \hline
        \multirow{3}{*}{15} 
            ~ & 0    & 0.54    & 0.10 & 0.82    & 0.12 & 0.84    & 0.00 & 0.00 & 0.01 & 0.77    & 0.00 & 0.00 & 0.01 & 0.79 \\
            ~ & 1    & 0.87    & 0.17 & 0.96    & 0.22 & 0.97    & 0.00 & 0.00 & 0.03 & 0.94    & 0.00 & 0.00 & 0.05 & 0.95 \\
            ~ & 9    & 0.60    & 0.17 & 0.91    & 0.12 & 0.94    & 0.00 & 0.00 & 0.02 & 0.62    & 0.00 & 0.00 & 0.01 & 0.76 \\
        \hline
    
    \end{tabular}
    \label{tab:upsample_combined}
\vskip -0.2in
\end{table*}
\fi

%% file: sections/related.tex
\section{Related Work}\label{sec:relwork}

Given the distributed training process in
FL, poisoning attacks
represent an even larger threat than in traditional ML systems. 
For instance,  poisoning availability attacks have been shown effective in Federated Learning in recent work~\cite{fangLocalModelPoisoning2020a,Shejwalkar2021ManipulatingTB}. Targeted model poisoning attacks impact a small population, or introduce a backdoor in the models to mis-classify instances containing the trigger~\cite{bhagojiModelPoisoningAttacks2018,bagdasaryan2020backdoor,sun2019can,wangAttackTailsYes2020}. 
Researchers also proposed methods to defend the FL protocol from adversaries, such as Byzantine-resilient or trust-based aggregation rules~\cite{blanchardMachineLearningAdversaries2017, mhamdiHiddenVulnerabilityDistributed2018, yinByzantineRobustDistributedLearning2018a, alistarhByzantineStochasticGradient2018, guerraoui2021garfield, sun2021flwbc, cao2021fltrust, mhamdi2021collaborative, mhamdi2021distributed}. 
However, \cite{fangLocalModelPoisoning2020a} and \cite{Shejwalkar2021ManipulatingTB} systematically analyzed Byzantine-robust aggregation schemes, showing that an adversary controlling compromised clients can bypass these defenses with poisoning availability attacks.
Also, targeted poisoning attacks can bypass Byzantine-resilient aggregation, such as Krum~\cite{bagdasaryan2020backdoor}. 
Methods protecting from model poisoning include filtering of malicious gradients for availability attacks~\cite{fangLocalModelPoisoning2020a,Shejwalkar2021ManipulatingTB,signguard2021} and gradient clipping for targeted attacks~\cite{sun2019can, bagdasaryan2020backdoor}.

%% file: sections/conclusion.tex
\section{Conclusion}
\label{sec:conclusion}

We examined the effects that a network-level adversary can have on the final model's accuracy on a target population in cross-device FL.
We proposed a new attack, based on a procedure for identifying the set of clients mostly contributing towards a target class of interest to the adversary, and showed that it can be amplified by coordinated model poisoning attacks.
We showed that our attacks are effective for realistic scenarios where the communication is encrypted and the attacker
has limited network visibility.
We also explored defensive approaches, and found that our  UpSampling  mechanism can be extremely successful when paired with clipping of model updates, against very powerful network-level attacks.


\section*{Acknowledgements}

This research was supported by the Department of Defense Multidisciplinary Research Program of the University Research Initiative (MURI) under contract W911NF-21-1-0322.